\documentclass[final,5p,times,twocolumn,a4paper]{elsarticle}

\usepackage{amsfonts}
\usepackage{amsmath}
\usepackage{amssymb}
\usepackage{graphicx}

\journal{Optics Communications}

\begin{document}

\newcommand{\intl}{\int_{-\infty}^{+\infty}}
\newcommand{\disp}{\displaystyle}
\newcommand{\Real}{\mathrm{Re}}
\newcommand{\Imag}{\mathrm{Im}}
\def\sech{{\rm sech}}

\begin{frontmatter}

\title{Soliton strings  and interactions in mode-locked lasers}

\author[cu]{Mark J. Ablowitz}
\ead{mark.ablowitz@colorado.edu}

\author[tph]{Theodoros P. Horikis\corref{cor1}}
\ead{horikis@uop.gr}

\author[cu]{Sean D. Nixon}
\ead{sean.nixon@colorado.edu}

\address[cu]{Department of Applied Mathematics, University of
Colorado, 526 UCB, Boulder, CO 80309-0526}

\address[tph]{Department of Computer Science and Technology, University of
Peloponnese, Tripolis 22100, Greece}

\cortext[cor1]{Corresponding author}

\begin{abstract}
Soliton strings  in mode-locked lasers are obtained using a variant of the
nonlinear Schr\"{o}dinger equation, appropriately modified to model power
(intensity) and energy saturation. This equation goes beyond the well-known master
equation often used to model these systems. It admits mode-locking and soliton
strings in both the constant dispersion and dispersion-managed systems in the (net)
anomalous and normal regimes; the master equation is contained as a limiting case.
Analysis of soliton interactions show that  soliton strings  can form when pulses
are a certain distance apart relative to their width. Anti-symmtetric bi-soltion
states are also obtained. Initial states mode-lock to these states under evolution.
In the anomalous regime individual soliton pulses are well approximated by the
solutions of the unperturbed nonlinear Schr\"{o}dinger equation, while in the
normal regime the pulses are much wider and strongly chirped.
\end{abstract}
\end{frontmatter}

\section{Introduction}
Ultra-short pulses in mode-locked (ML) lasers  are  the topic of extensive research
due to their wide range of applications ranging from communications \cite{haus1},
to optical clock technology \cite{cundiff-book} and even to measurements of the
fundamental constants of nature \cite{fischer}. Although ML lasers have been
studied for many years \cite{lamb,haus1} it is only recently that researchers have
begun to fully understand and explore their complicated dynamics. Here we discuss
pulses associated with a model equation  which has both power (intensity) and
energy saturation. This  equation  goes beyond the well-known master equation
\cite{master3,haus1} which is contained in the limiting case of small power. The
master equation has gain and filtering terms which are saturated by energy;  there
is an additional nonlinear gain/loss term which grows with power (intensity). On
the other hand, this model critically differs by having the loss term saturated by
power (intensity). We show below that this equation admits soliton strings,
including anti-symmetric bi-soliton states, in both the anomalous and normal
regimes and these modes can  mode-lock from initial states under evolution, i.e. we
find these soliton states to be attractors. These results are consistent with
observations from recent experiments in both the anomalous \cite{tang,grelu} and
normal \cite{chong2} regimes where in the latter case higher-order   anti-symmetric
solitons (or bi-solitons) were observed.

A simplified equation without gain/loss terms which is sometimes used to model
ultrashort pulses in mode-locked (ML) lasers is the ``classical'' nonlinear
Schr\"{o}dinger (NLS) equation. The NLS equation in the anomalous regime exhibits
both fundamental and high-order soliton solutions. While the fundamental solutions
(single solitons) are stationary the higher states exhibit nontrivial evolution. In
the normal regime the NLS equation does not have localized decaying solutions.
Indeed, the NLS equation has only dark solitons, namely states that do not decay at
infinity but form on a nontrivial background. On the other hand to properly
describe ML lasers, in addition to chromatic dispersion and Kerr nonlinearity
(self-phase modulation), saturable gain, filtering and intensity discrimination
should be taken into account. In this regard, the so-called master equation
\cite{master3,haus1} is  an important extension of the NLS equation modified
accordingly to contain gain and filtering saturated by energy (the time integral of
the pulse power), while there is an additional (sign dependent) cubic nonlinear
term that takes into account additional gain/loss. The master equation has only a
small parameter regime where mode-locking to stable soliton states occurs
\cite{kutz};   it has not been shown to support  higher-order states. In fact, for
certain values of the parameters this equation exhibits a range of phenomena
including: mode-locking evolution, pulses which disperse into radiation, and some
whose amplitude grows rapidly \cite{kutz2}. In the latter case, if the nonlinear
gain is too high, the linear attenuation terms are unable to prevent the pulse from
blowing up, suggesting the breakdown of the master mode-locking model \cite{kutz}.
Thus an improved model is desirable.

If the pulse energy is taken to be constant the master equation reduces to a cubic
nonlinear Ginzburg-Landau (GL) type system. Such GL systems in the anomalous regime
have been found to support steady high-order soliton solutions
\cite{akhmediev_trains}; they also contain a wide range of solutions including
unstable, chaotic and quasi-periodic states and even blow-up can occur.
Furthermore, soliton interaction within the framework of such GL systems is
complicated. It is found \cite{akhmediev_trains} that both in-phase and anti-phase
high-order soliton  states are either unstable or weakly stable. In either case
these states are not attractors \cite{akhmediev3}. Hence they do not correspond to
observations of higher order soliton states in a ML laser system which has
discrete, nearly  fixed separations \cite{tang}. Furthermore, these GL equations do
not exhibit stable pulses in the normal regime \cite{grelu}.

The distributive both power (intensity) and energy saturation model which we refer
to as the power-energy saturation (PES) equation is given in nondimensional form as
\begin{gather}
i\psi_z+\frac{d(z)}{2}\psi_{tt}+n(z)|\psi|^2\psi= \nonumber
\\ \frac{ig}{1+\epsilon E}\psi+ \frac{i\tau}{1+\epsilon E}\psi_{tt} -
\frac{il}{1+\delta P}\psi
\label{PES}
\end{gather}
where $\psi(z,t)$ is the slowly varying  electromagnetic pulse envelope,
$E(z)=\intl |\psi|^2 dt$ is the pulse energy, $P(z,t)=|\psi|^2$ is the
instantaneous pulse power and the parameters $g$, $\tau$, $l$, $\epsilon$, $\delta$
are all positive, real constants.  The dispersion is represented by $d(z)$, which
is taken to either be constant ($d(z)=d_0>0$ corresponds to the anomalous regime
while $d(z)=d_0<0$ represents normal) or a dispersion-managed system (see Section
\ref{dm.section}). The dimensionless nonlinear coefficient will be taken to be
unity ($n(z)=1$) in the first part of the article and varying between zero and one
in the dispersion-managed case, as holds in many experiments
\cite{Ildaypsat,quraishi}. The first term on the right hand side represents
gain-saturated by energy, the second is spectral filtering-saturated by energy and
the third is loss saturated by power (intensity). The coefficients  $\epsilon$,
$\delta$ are related to the saturation energy and power respectively. If we expand
the denominator of the loss term in a Taylor series in the limit of small power and
keep only the first two terms, we reduce to the master-equation! Furthermore, if
one expands the power saturation term in Eq. (\ref{PES}) and keep higher order
nonlinearities (e.g. fifth order nonlinearities) higher nonlinear GL systems
result. The energy and power saturation terms in Eq. (\ref{PES}) are essentially
the simplest such forms (inversely linear in their saturation effects). This type
of power saturated loss has been used in lumped models to describe experimental ML
pulses \cite{Ildaypsat,chong2} and is widely used as a saturable loss model (e.g.
semiconductor saturable loss mirrors--SeSAMs). In \cite{horikis,horikis2} the
distributive model was employed to show that single soliton states model-lock over
a large region of parameter space. More recently it was shown \cite{Kartner} that
dispersion-managed models  with power (intensity)-energy saturation are in good
agreement with experimental results in mode-locked Ti:sapphire lasers. It is
therefore important to study the above distributed model.

Dimensional values associated with a typical Ti:Sapphire laser system are
\cite{ilday2}: $\beta''=60\mathrm{fs}^2/\mathrm{mm}$ is the group velocity
dispersion, $\gamma= 1 \mathrm{/cmMW}$ the nonlinear coefficient $P_*=5\mathrm{MW}$
the characteristic power, $E_*=P_*t_*$ the characteristic energy,
$z_*=2\mathrm{mm}$ the corresponding characteristic-nonlinear length ($z_*=
1/\gamma P_*$), $t_*=10\mathrm{fs}$ the characteristic time, $g_*=20\mathrm{dB}$
the dimensional gain, $l_*=0.15 \mathrm{1/mm}$ the dimensional distributive
saturable loss parameter, $\Delta\omega=100$ THZ the frequency cutoff,
$E_{\mathrm{sat}}=10\mathrm{nJ}$ and $P_{\mathrm{sat}}=2\mathrm{MW}$ are the
saturated energy and power, respectively. Hence: $d_0=\beta'' z_*/t_*^2$,
$g=g_*z_*$, $\tau=g_*z_*/\Delta\omega^2 t_*^2$, $\epsilon=E_*/E_{\mathrm{sat}}$ and
$\delta=P_*/P_{\mathrm{sat}}$. These dimensional values lead to phenomena which are
consistent  with the results presented in this article.

\section{Constant dispersion systems}

The effects of energy and power saturation are crucial in both the anomalous and
normal regimes. From ML laser behavior, the gain and filtering mechanisms are
related to the energy of the pulse, while the loss is related to the power
(intensity) of the pulse.  Indeed, passive mode-locking generally utilizes
saturable absorbers which are modeled  here by a distributive loss term saturated
with power.

In the constant anomalous regime $(d(z)=d_0>0)$ the saturation terms help localize
an intense pulse preventing it from reaching a singular state; i.e. ``infinite"
energy or a blow-up in amplitude does not result. Indeed, if blow-up were to occur
that would mean that both the amplitude and the energy of the pulse are large,
hence, the perturbing effects are very small thus reducing Eq. (\ref{PES}) to the
unperturbed NLS equation, which admits a stable, finite solution. In addition, when
localization occurs, the perturbing effects are small essentially reducing the
equation to the unperturbed NLS equation. Indeed, it has been shown \cite{horikis2}
that when mode-locking occurs the resulting individual pulses are essentially
solitons of the unperturbed NLS equation; the perturbing influence of these terms
yields the mode-locking mechanism.

In the constant normal regime $(d(z)=d_0<0)$, three regimes are observed
\cite{AHnormal}: (a) when the loss is much greater than the gain the pulse decays
to zero, (b) when the loss is again the prominent effect but sufficient gain exists
in the system to sustain a very slowly decaying evolution resulting in a
``quasi-soliton'' state and (c) the soliton regime above a certain value of gain.
This soliton differs from its anomalous dispersion counterpart in that it does not
obey the unperturbed NLS equation, it is much wider and strongly chirped.

\begin{figure}[!htbp]
    \centering
    \includegraphics[width=2.5in]{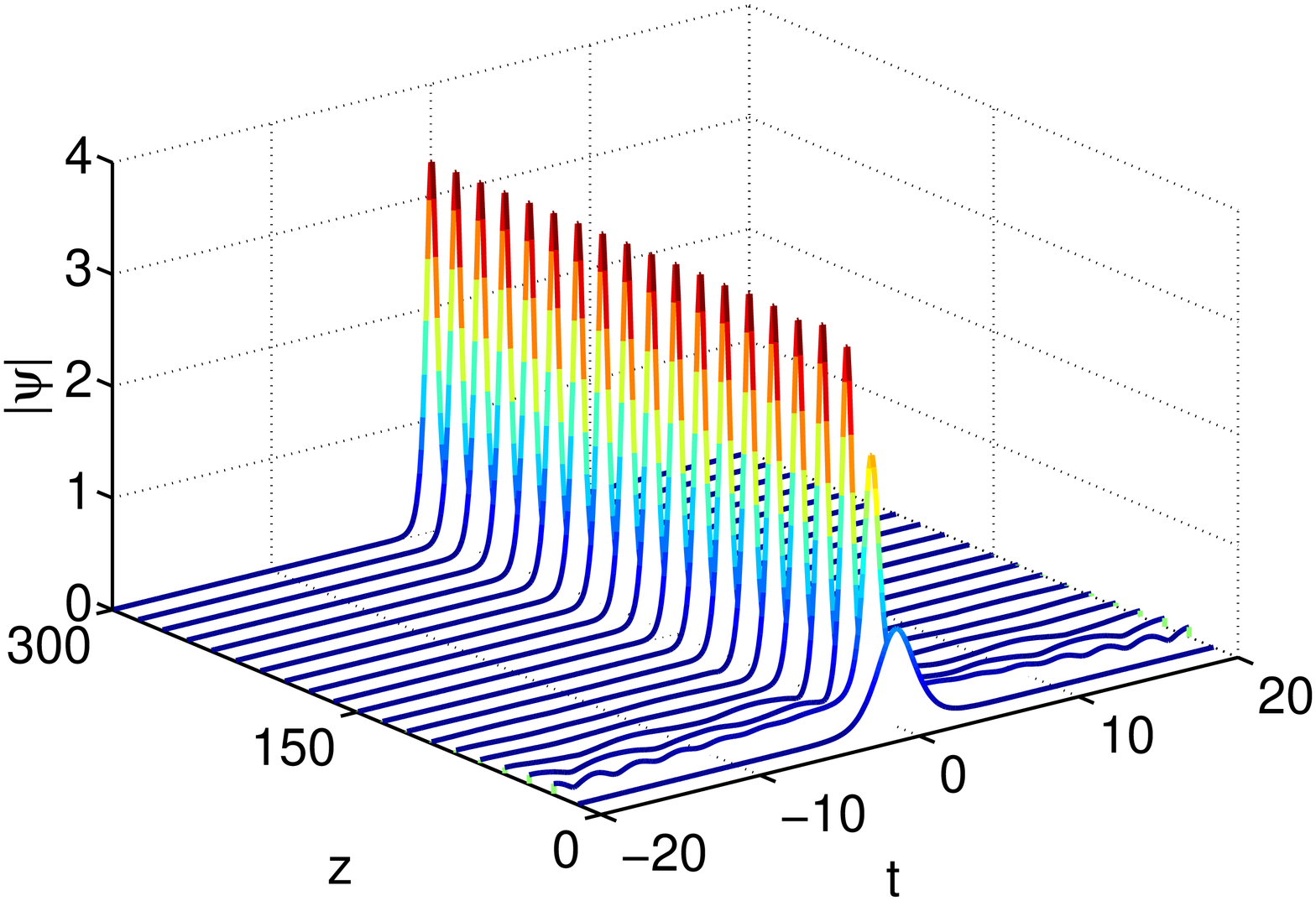}
    \includegraphics[width=2.5in]{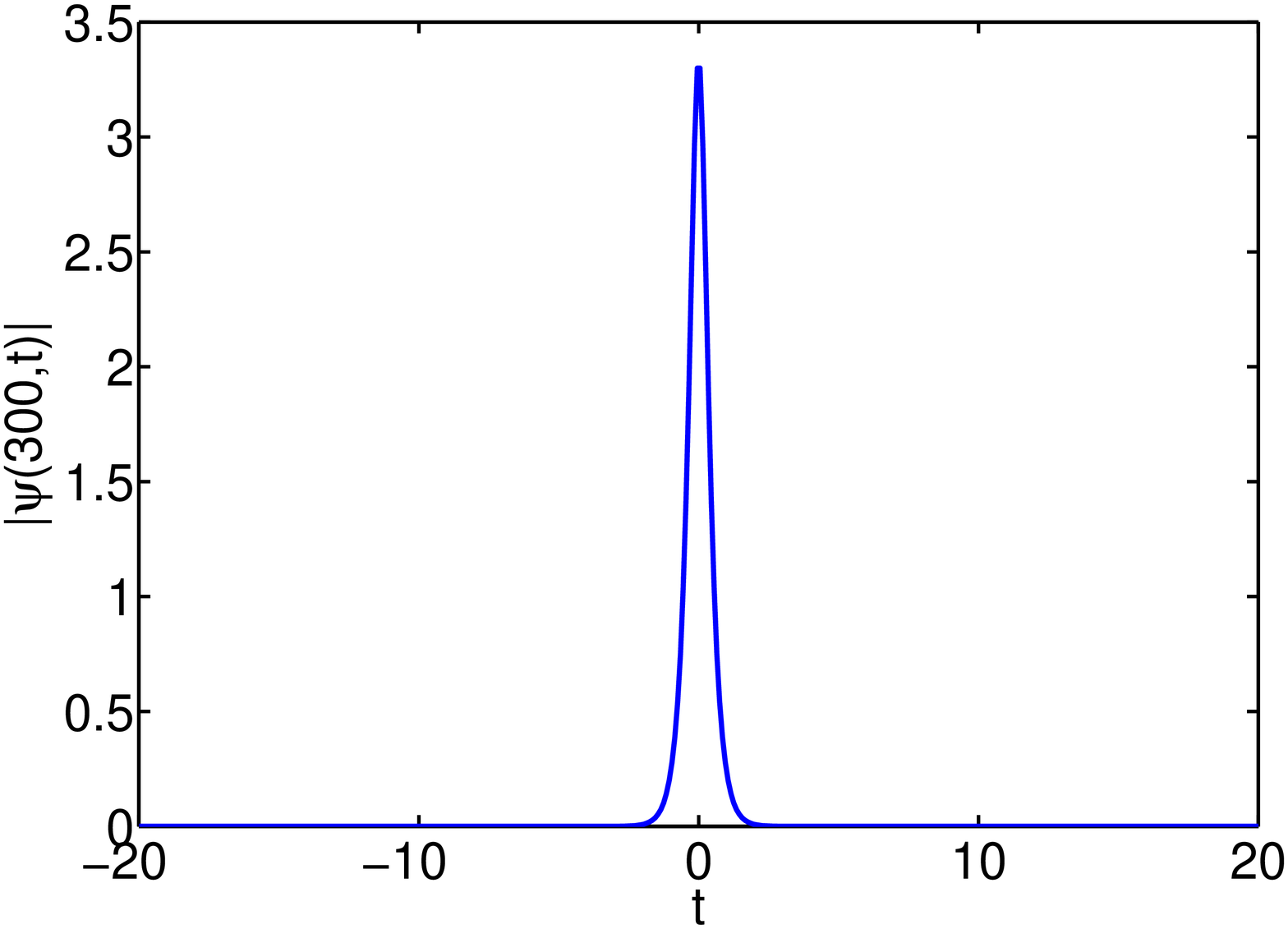}\\
    \includegraphics[width=2.5in]{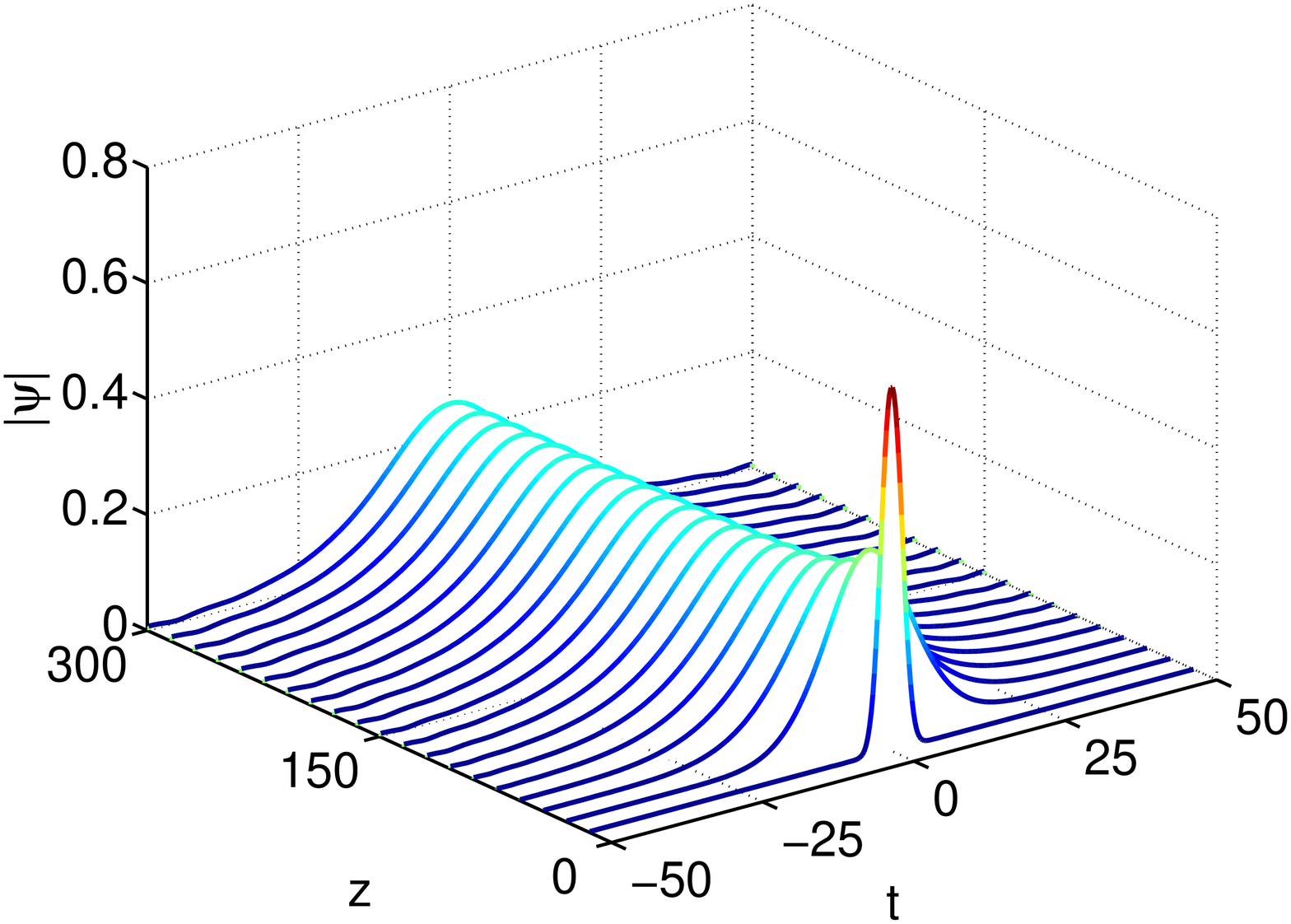}
    \includegraphics[width=2.5in]{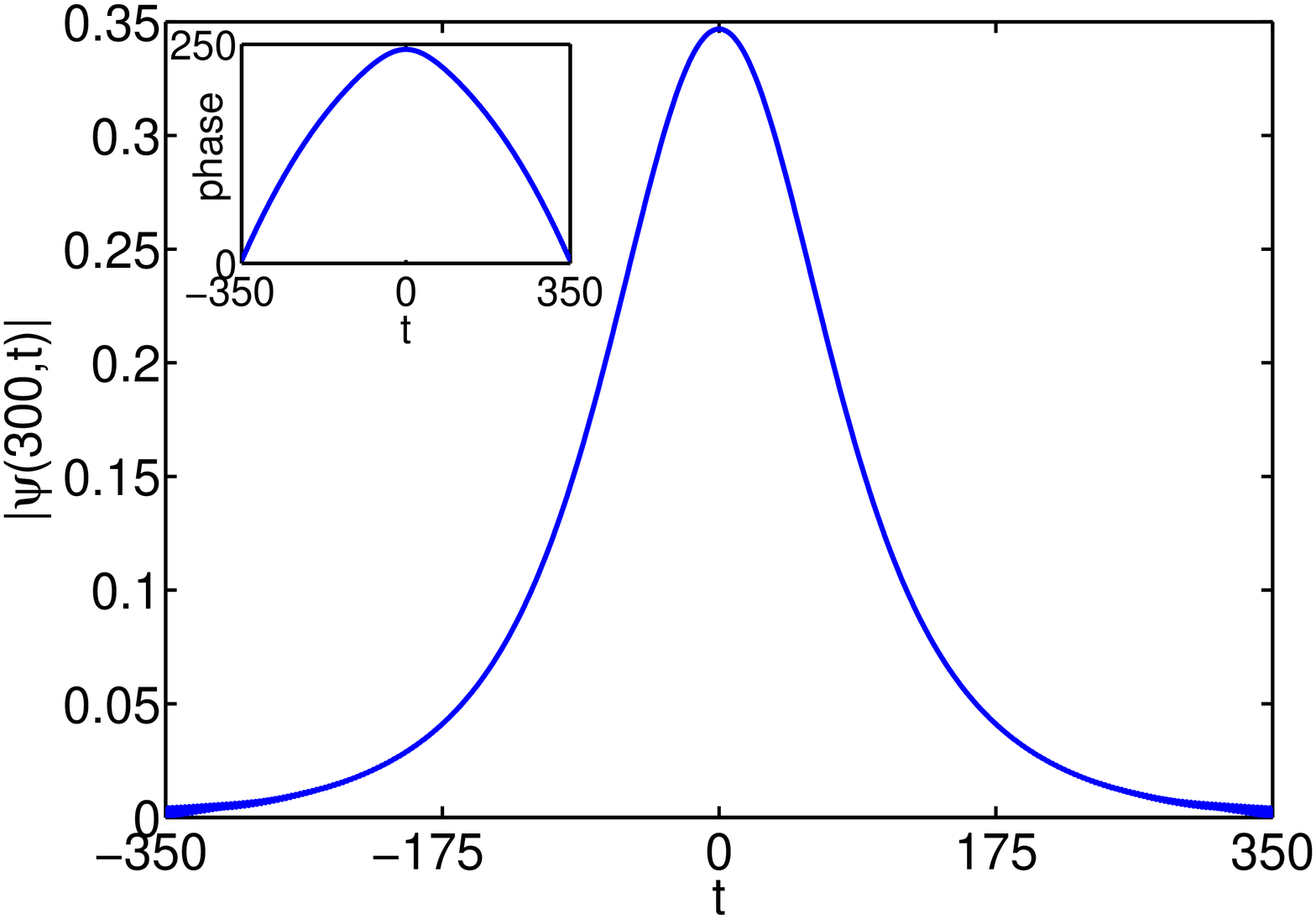}
    \caption{(Color online) Mode-locking evolution for a single soliton evolving from a unit gaussian
    initial state in the anomalous (top two figures) and normal (bottom two figures) regimes; states at $z=300$ are
    depicted below each evolution.  For the normal regime the phase-chirp is shown in the inset.}
    \label{single.fig}
\end{figure}

For example in Fig. \ref{single.fig} a soliton mode is depicted which evolves from
a unit gaussian state in the anomalous ($d_0=1$) and normal ($d_0=-1$) regimes
(here  $\tau=l=0.1$, $\epsilon=\delta=1$ and $g=0.5$ in the anomalous case, $g=1.5$
in the normal case)

\begin{figure}[!htbp]
    \centering
    \includegraphics[width=2.5in]{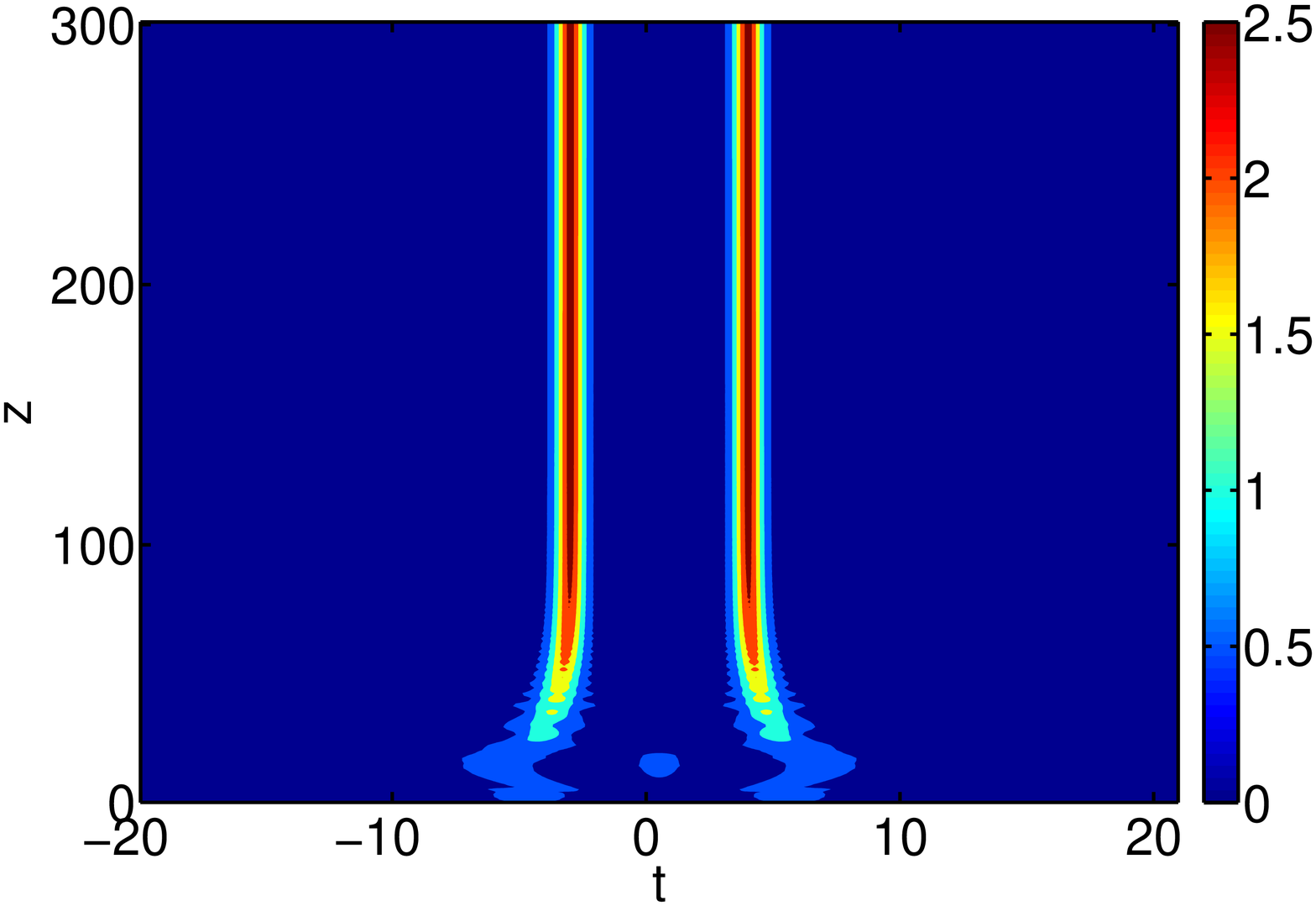}
    \includegraphics[width=2.5in]{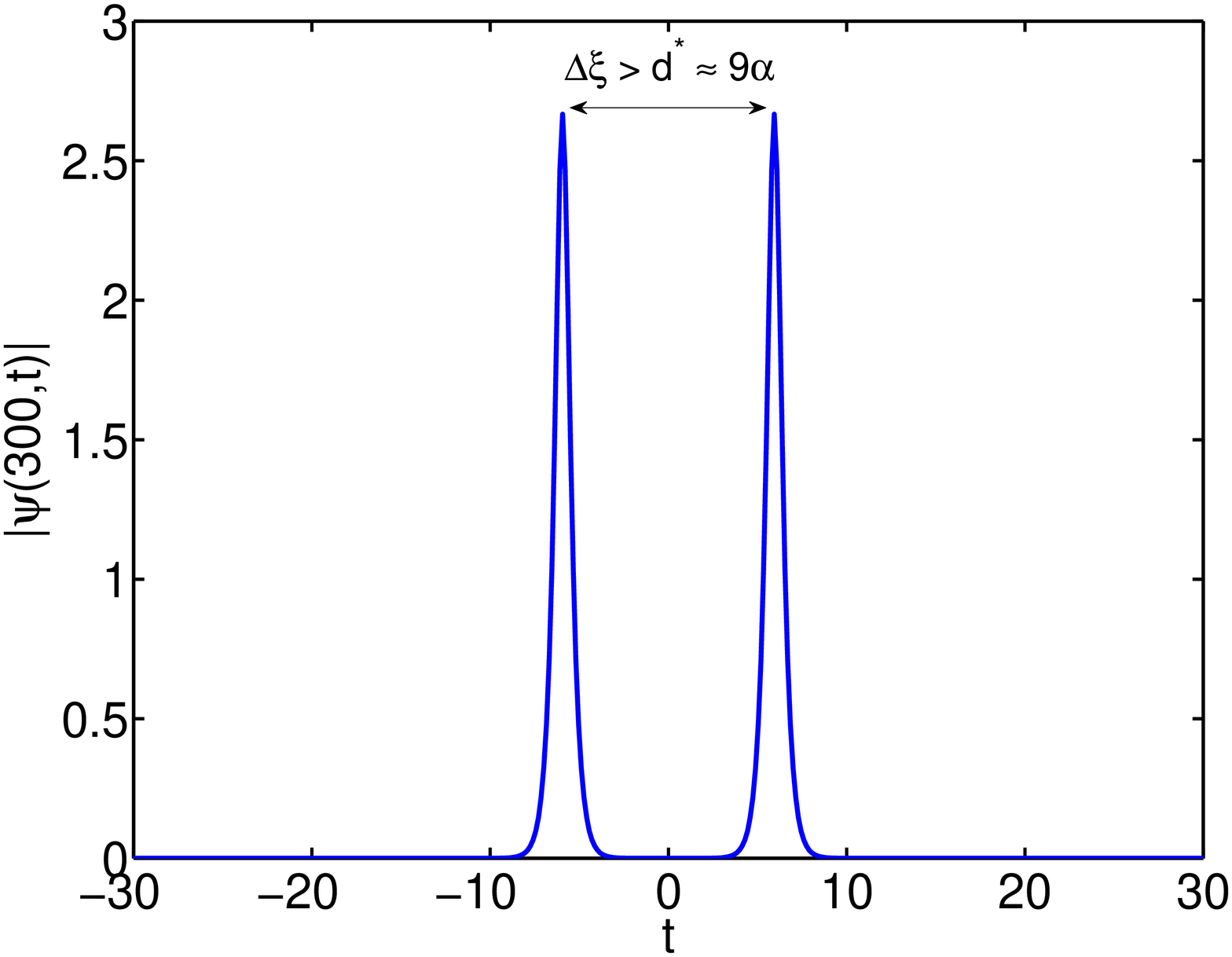}
    \caption{(Color online) Mode-locking evolution for an in phase two soliton state of
    the anomalously dispersive PES (top) equation and the
    resulting soliton profile (bottom) at $z=300$. Here $g=0.5$. }
    \label{2sol.fig}
\end{figure}

High-order solitons of the anomalous regime $(d_0=1>0)$ are discussed first. Here
solitons are obtained when the gain is above a certain critical value, $g>l$,
otherwise pulses dissipate and eventually vanish \cite{horikis2}.  As gain becomes
stronger additional soliton   states are possible and 2, 3, 4 or more coupled
pulses are found to be supported. Here we set $\tau=l=0.1$, $\epsilon=\delta=1$ and
vary the gain parameter $g$.  The value of $\Delta \xi/\alpha$, where $\Delta \xi$
and $\alpha$ are the pulse separation and pulse width respectively, is an important
parameter. The full width of half maximum (FWHM) for pulse width is used, $\Delta
\xi,$  is measured between peak values of two neighboring pulses and $\Delta \phi$
is the phase difference between the peak amplitudes. With sufficient gain $(g=0.5)$
Eq. (\ref{PES}) is evolved starting from unit gaussians with initial peak
separation $\Delta\xi = 10$ and $\Delta\xi/\alpha=8.5$.  The evolution and final
state are depicted in Fig. \ref{2sol.fig}.
For the final state we find $\Delta\xi/\alpha \approx 10$.  The resulting
individual pulses are similar to the single soliton mode-locking case
\cite{horikis2}, i.e. individual pulses  are approximately solutions of the
unperturbed NLS equation, namely hyperbolic secants. The pulses differ from a
single soliton in that the individual pulse energy is smaller then that observed
for the single soliton mode-locking case for the same choice of $g$, while the
total energy of the two soliton state is higher.  This is due to the non-locality
of energy saturation in the gain and filtering terms.

To investigate the minimum distance, $d^*$, between the solitons in order for no
interactions to occur we evolve  equation starting with two solitons. If the
initial two pulses are sufficiently far apart then the propagation evolves to  a
two soliton state and the resulting pulses have a constant phase difference.  If
the distance between them is less than a critical value then the two pulses
interact in a way characterized by the difference in phase between the peaks
amplitudes: $\Delta \phi$. When initial conditions are symmetric (in phase) two
pulses are found to merge into a single soliton of Eq. (\ref{PES}). When the
initial conditions are anti-symmetric (out of phase by $\pi$) then they repel each
other until their separation is above this critical distance while retaining the
difference in phase, resulting in an  effective two pulse high-order soliton state.
This does  not occur in the pure NLS equation as shown in the perturbation section
below.

\begin{figure}[!htbp]
    \centering
    \includegraphics[width=2.5in]{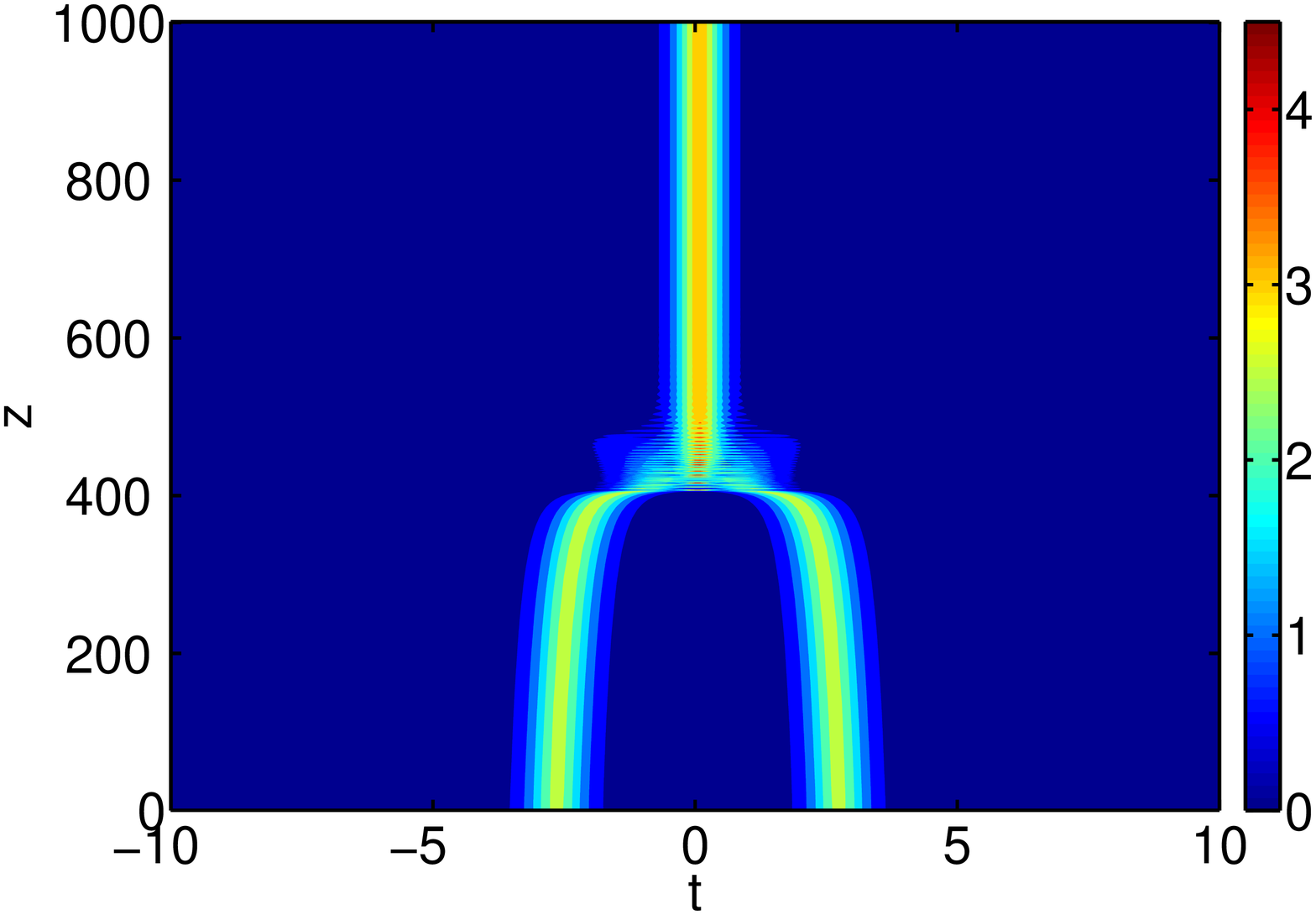}
    \includegraphics[width=2.5in]{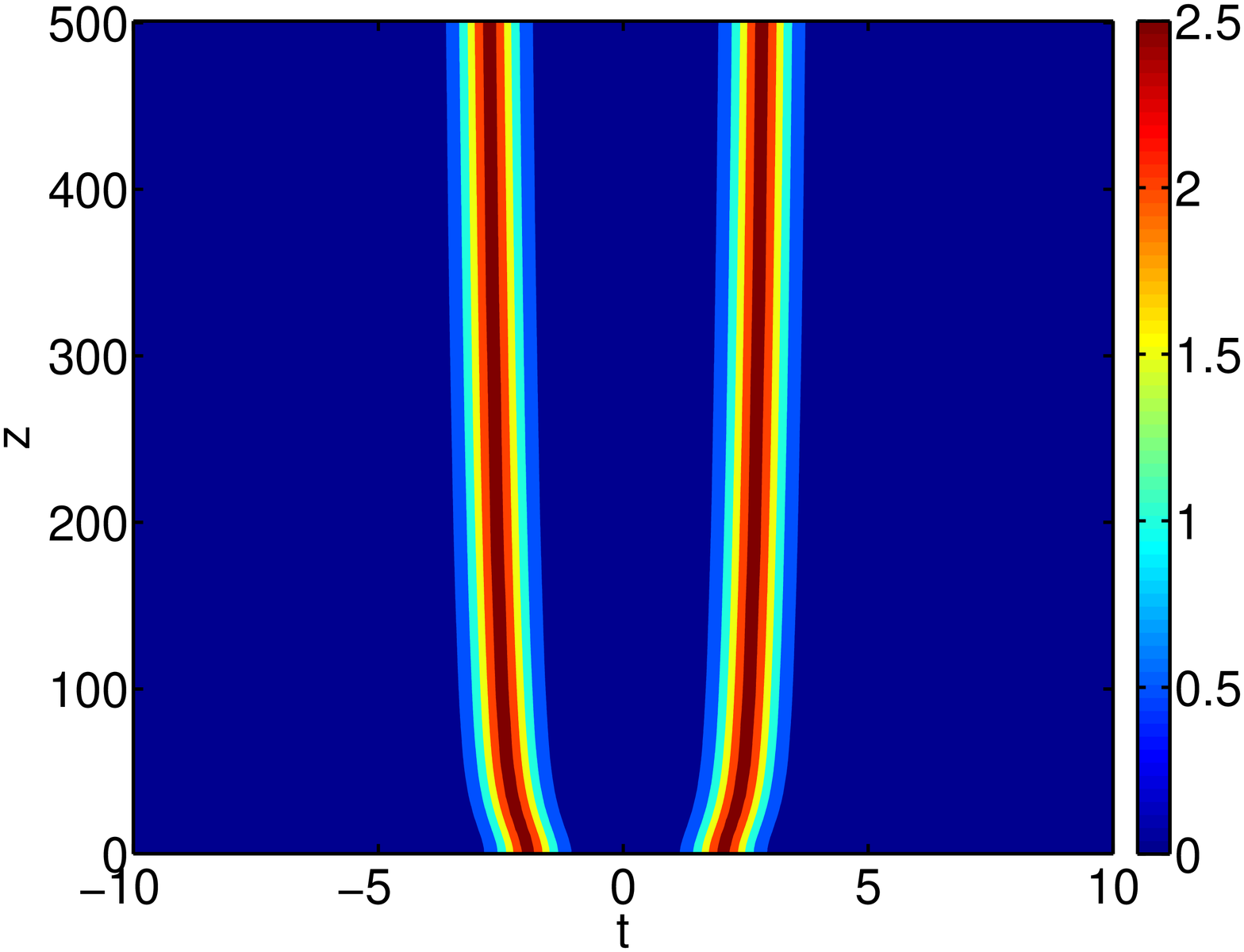}
    \caption{(Color online) Two pulse interaction when $\Delta\xi<d^*$. Initial pulses
    $(z=0)$ in phase: $\Delta \phi=0$,
    $\Delta\xi/\alpha \approx 7$ (top) merge while those out of phase by
    $\Delta\phi=\pi$ with $\Delta\xi/\alpha \approx 6$ (bottom) repel. Here $g=0.5$.}
    \label{2contours}
\end{figure}
In the constant dispersion case this critical distance is found to be $\Delta
\xi=d^*\approx 9\alpha$ (see Fig. \ref{2sol.fig}) corresponding to soliton initial
conditions. Interestingly, this is consistent with the experimental observations of
Ref. \cite{tang}. To further illustrate, we plot the evolution of these cases in
Fig. \ref{2contours}.  At $z=500$ for the repelling solitons $\Delta\xi/\alpha=
8.9$.

\section{Perturbation theory}

To obtain a more fundamental understanding of these numerical results we use
perturbation theory to find evolution equations for the six parameters which define
the interactions. These are: the pulse heights and velocities $\eta_k$ and $V_k$
respectively for $k=1,2$, the distance between pulse peaks $\Delta \xi$, and the
difference in phase $\Delta \phi$.  We take the subscripts 1 and 2 to denote the
left and right pulse respectively and $\Delta \phi$ to be the phase of the right
pulse minus that of the left pulse.  Similarly, $\Delta \eta = \eta_2 - \eta_1$ and
$\Delta V =  V_2 - V_1$.  Since the individual pulses are well approximated by
solitons of the unperturbed NLS equation, the pulses are considered to be solutions
of this equation which vary slowly under the perturbing effects of gain, loss, and
filtering.

Let us denote the NLS equation with perturbation $F[\psi]$ as
\begin{eqnarray}
   i\psi_z + \frac{1}{2}\psi_{tt} + \left|\psi\right|^2\psi = F[\psi]
   \label{pertNLS}
\end{eqnarray}
where
\begin{eqnarray*}
F[\psi] = \frac{ig}{1+\epsilon E}\psi+ \frac{i\tau}{1+\epsilon E}\psi_{tt} -
\frac{il}{1+\delta P}\psi
\end{eqnarray*}
We write the four parameter family of soliton solutions to the unperturbed or
classical NLS equation ($F=0$) as follows
\begin{subequations}
\begin{gather}
\psi= ue^{i\phi},\quad u=\eta \sech (\eta\theta)\\
\xi = \int_{0}^z V dz + \xi_0,\quad \theta = t - \xi,\\ \sigma = \displaystyle\int
_{0}^{z} (\mu + \frac{V^2}{2})dz  + \sigma_0,\quad \phi = V\theta + \sigma
\end{gather}
\label{forma}
\end{subequations}
where $\mu$, $V$, $\xi_0$, $\sigma_0$  define height/width, speed, temporal shift
and phase shift of the soliton respectively. For the classical NLS equation without
gain, filtering and loss ($F = 0$), $\mu$, $V$, $\xi_0$, $\sigma_0$ are constant.
We also note that $\eta$ is directly related to $\mu$ by $\mu = \eta^2/2$.  The
effect of the right-hand-side being nontrivial is approximated by allowing $\mu$,
$V$, $\xi_0$, $\sigma_0$ to vary slowly in $z$. The evolution of these parameters
is found by satisfying the following integral conservation laws modified by
$F[\psi]$ and derived from Eq. (\ref{pertNLS})
\begin{subequations}
\begin{align}
\frac{d}{dz}\int\left|\psi\right|^2&= 2\Imag\int F[\psi]\psi^*\\
\frac{d}{dz}\Imag\int \psi \psi_t^*&=2\Real\int F[\psi] \psi_t^*\\
\frac{d}{dz}\int t \left|\psi\right|^2&=- d_0\Imag \int \psi \psi^*_t
+ 2 \Imag \int t F[\psi]\psi^*\\
\Imag\int \psi_z \psi_{\mu}^*&=- \frac{d_0}{2}\Real\int \psi_t \psi_{t\mu}^*
 +\Real\int  \left| \psi\right|^2 \psi \psi_{\mu}^* \nonumber\\ &\hspace{2.7cm} - \Real \int F[\psi]\psi_{\mu}^*
\end{align}
\label{intrel}
\end{subequations}
where all integrals are taken over $-\infty<t<\infty$.

Along with the effects of gain, loss, and filtering, the effect of tail interaction
between the two solitons is also treated as a perturbation. Since solitons are
widely separated to leading order the full solution may be viewed as the
superposition of two single solitons $\psi = \psi_1 + \psi_2$; we take $\psi_1$ to
be the soliton on the left and $\psi_2$ to be the soliton on the right.  When
$\psi_1$ is locally the dominant term $\psi_2$ is treated as a small parameter and
we expand the nonlinear term about $\psi_1.$  If for the moment we omit the
$F[\psi]$ term  the evolution of $\psi_1$ is well approximated by
\begin{align*}
i\psi_{1z} + \frac{1}{2} \psi_{1tt} +  \left|\psi_1\right|^2\psi_1 &= - 2\left|\psi_1\right|^2\psi_2 - \psi_1^2\
\psi_2^*\\ &= G[\psi_1,\psi_2]
\end{align*}
and similarly an equation for when $\psi_2$ dominates is found to satisfy
\begin{align*}
i\psi_{2z} + \frac{1}{2} \psi_{2tt} +  \left|\psi_2\right|^2\psi_2 &= - 2\left|\psi_2\right|^2\psi_1 - \psi_2^2\
\psi_1^*\\&=G[\psi_2,\psi_1]
\end{align*}
Notice that if $\psi_1$ and $\psi_2$ satisfy this system of coupled PDEs then their
sum  $\psi$ satisfies NLS.  Any contributions from interaction in the gain, loss,
and filtering terms would be a higher order term.  For a more thorough treatment of
this method for analyzing soliton interactions see Ref. \cite{KarpSolo} and for
unperturbed NLS see Ref. \cite{yijianke}.  Simplifications result if we assume that
$\Delta V = V_2-V_1$ is small so that from the definition $\Delta\phi \approx
-\bar{V}\Delta\xi + \Delta\sigma$.  We also assume $\Delta \eta = \eta_2-\eta_1$ is
small and approximate $\eta_1$ and $\eta_2$ as $\bar{\eta}$. For all variables the
bar denotes the mean over the two solitons and $\Delta$ the difference of the right
minus the left.  These assumptions are all satisfied in the numerical simulations.

Including the perturbing terms $F$ we have a system of equations coupled in their
perturbations as follows
\begin{eqnarray*}
i\psi_{1z} + \frac{1}{2} \psi_{1tt} +  \left|\psi_1\right|^2\psi_1 = G[\psi_1,\psi_2] + F[\psi_1] \\
i\psi_{2z} + \frac{1}{2} \psi_{2tt} +  \left|\psi_2\right|^2\psi_2 = G[\psi_2,\psi_1]  + F[\psi_2]
\end{eqnarray*}
both of which satisfy Eqs. (\ref{intrel}). Substituting the soliton ansatz for
$\psi _1$ and $\psi_2$ we find a system of first order differential equations for
the slowly varying parameters $\eta_k$, $V_k$, $\xi_{0k}$, and $\sigma_{0k}$.

\begin{figure}[!htbp]
    \centering
    \includegraphics[width=2.5in]{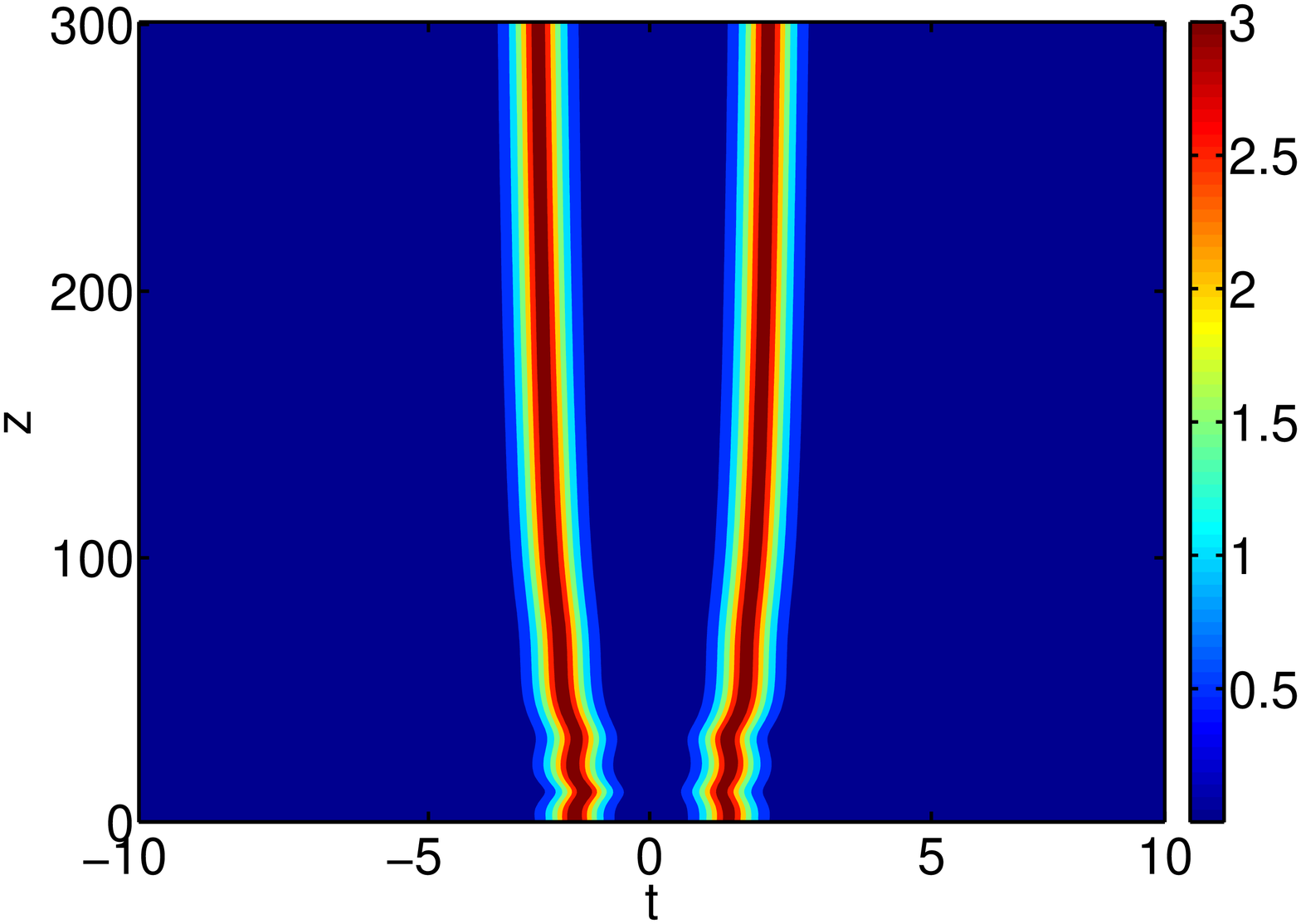}
    \includegraphics[width=2.5in]{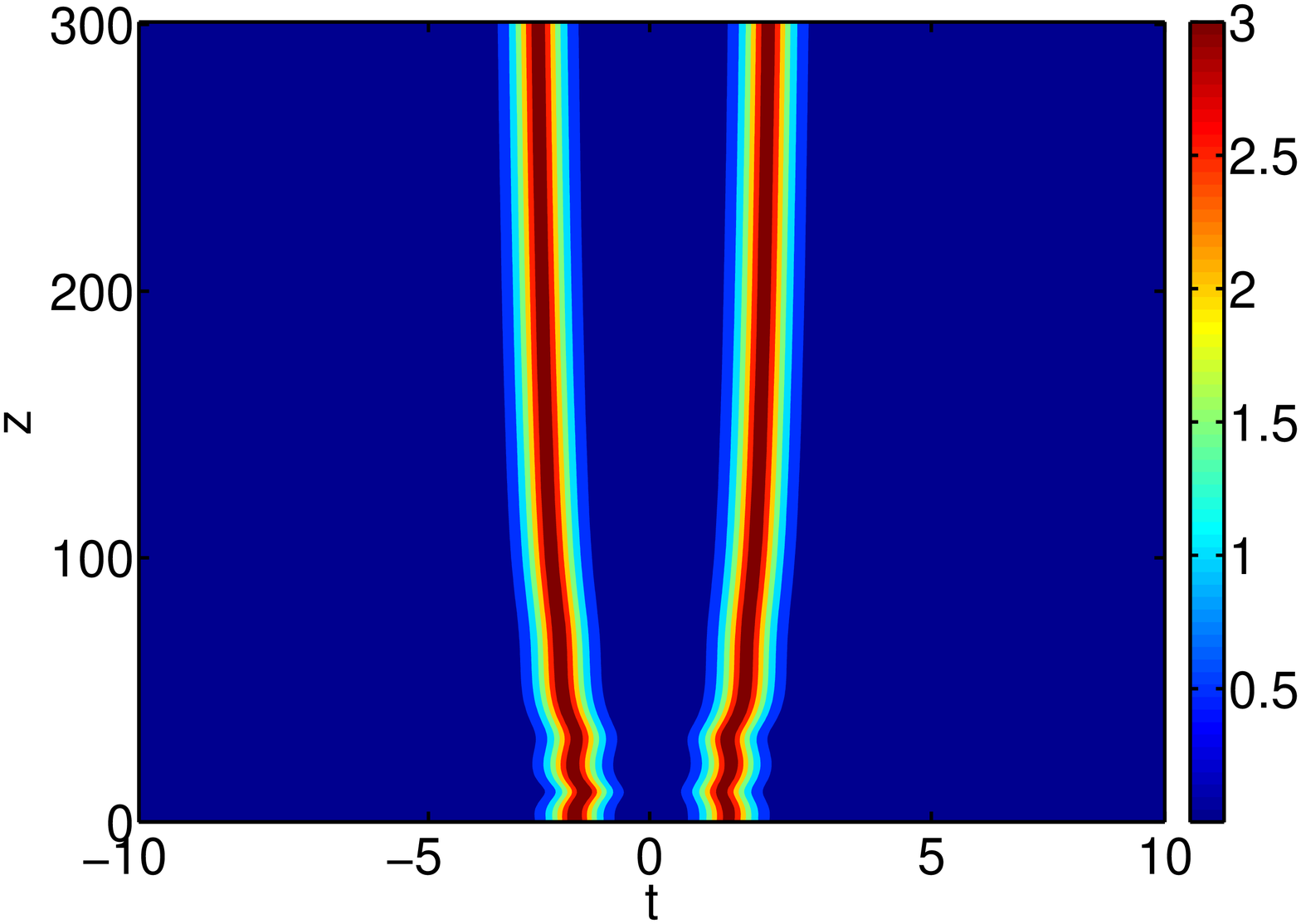}
    \includegraphics[width=2.5in]{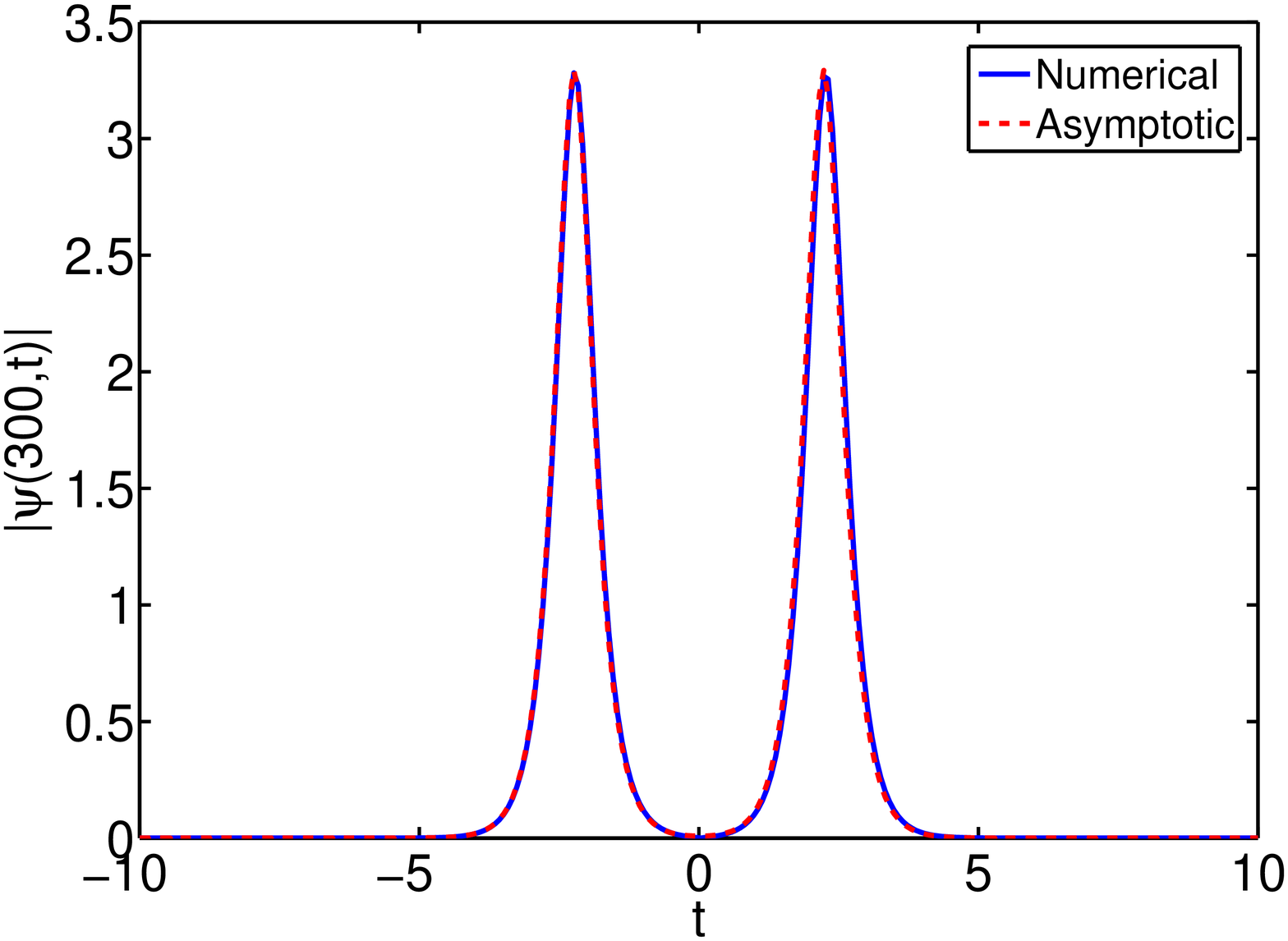}
    \caption{(Color online) Numerical simulation (top), asymptotic analysis (middle) and comparison of the numerical
    simulation and the asymptotic result at $z=300$ (bottom).  Here $g = 0.6$.}
    \label{compare}
\end{figure}

Subtracting the equations for $\frac{d\xi_{0k}}{dz}$, and $\frac{d\sigma_{0k}}{dz}$
we find
\[
\frac{d \Delta \xi_0}{dz} = 0,\quad  \frac{d \Delta \sigma_0}{dz} = 0
\]
which means that $\Delta \xi_{0}$, $\Delta \sigma_{0}$ are stationary.  Then, from
Eqs. (\ref{forma}) we have $\Delta\phi \approx  -\bar{V}\Delta\xi + \Delta\sigma$
and $\Delta\xi = \int_0^z (\Delta V)dz + \Delta \xi_0$ and by differentiating we
arrive at the following approximate evolution equations
\begin{align*}
	  \frac{d \Delta\xi }{dz}&= \Delta V + \Delta \xi_{0,z} = \Delta V\\
	\frac{d \Delta\phi}{dz}&=  -\bar{V}_z\Delta\xi -\bar{V}\Delta\xi_z + \Delta\sigma_z\\
	  &=     -\bar{V}\Delta V  +   (\bar{\eta}\Delta\eta + \bar{V}\Delta V )+\Delta \sigma_{0,z}
=\bar{\eta}\Delta\eta
	\end{align*}
The evolution equations are thus found to be
\begin{subequations}
\begin{align}
\frac{d\eta_k}{dz}  &= S_1(\eta_k,V_k) + (-1)^k 4\bar{\eta}^3 \sin(\Delta \phi)e^{-\bar{\eta} \Delta\xi}\\
\frac{dV_k}{dz}&=S_2(\eta_k,V_k) +  (-1)^{k+1}4\bar{\eta}^3 \cos(\Delta \phi)e^{-\bar{\eta} \Delta\xi}\\
\frac{d \Delta\xi }{dz}&=  \Delta V\\
\frac{d \Delta\phi}{dz}&= \bar{\eta}\Delta\eta
\end{align}
\label{evoeqn}
\end{subequations}
where
\begin{align*}
S_1(\eta,V, \bar{\eta}) &= g\left(\frac{2\eta}{E_0 +1} \right) - \tau \frac{1}{E_0 +1} \left( \frac{2}{3} \eta^3 +
2V^2 \eta \right) \\
&\hspace{2cm}+ l\left[2\eta \frac{1}{a-b} \log\left( \frac{a}{b}\right)\right] \\
S_2(\eta,V, \bar{\eta}) &= -\tau V\left( \frac{4}{3} \frac{\eta^2}{E_0+1} \right)
\end{align*}
Here $a$ and $b$ are the complex conjugate roots of the polynomial $x^2 +
2(1+2\eta^2)x+1=0$.  The energy for the two NLS solitons is $E_0=4\bar{\eta}$. Note
$a$ and $b$ can be chosen to be either root of the quadratic equation. A more
detailed derivation of these equations may be found in \cite{MLpert, nixon}, along
with an analysis of how interactions of solitons in Eq. (\ref{PES}) differ from
those in the unperturbed NLS equation. A comparison of the numerical results to
those found from the asymptotic analysis are seen in Fig. \ref{compare} for initial
conditions $\eta_{10} = \eta_{20} = 3.3$, $\Delta\xi_0 = 3$
($\Delta\xi_0/\alpha=6$), and $\Delta\phi_0 = 3\pi/4$.

\begin{figure}[!htbp]
    \centering
    \includegraphics[width=2.5in]{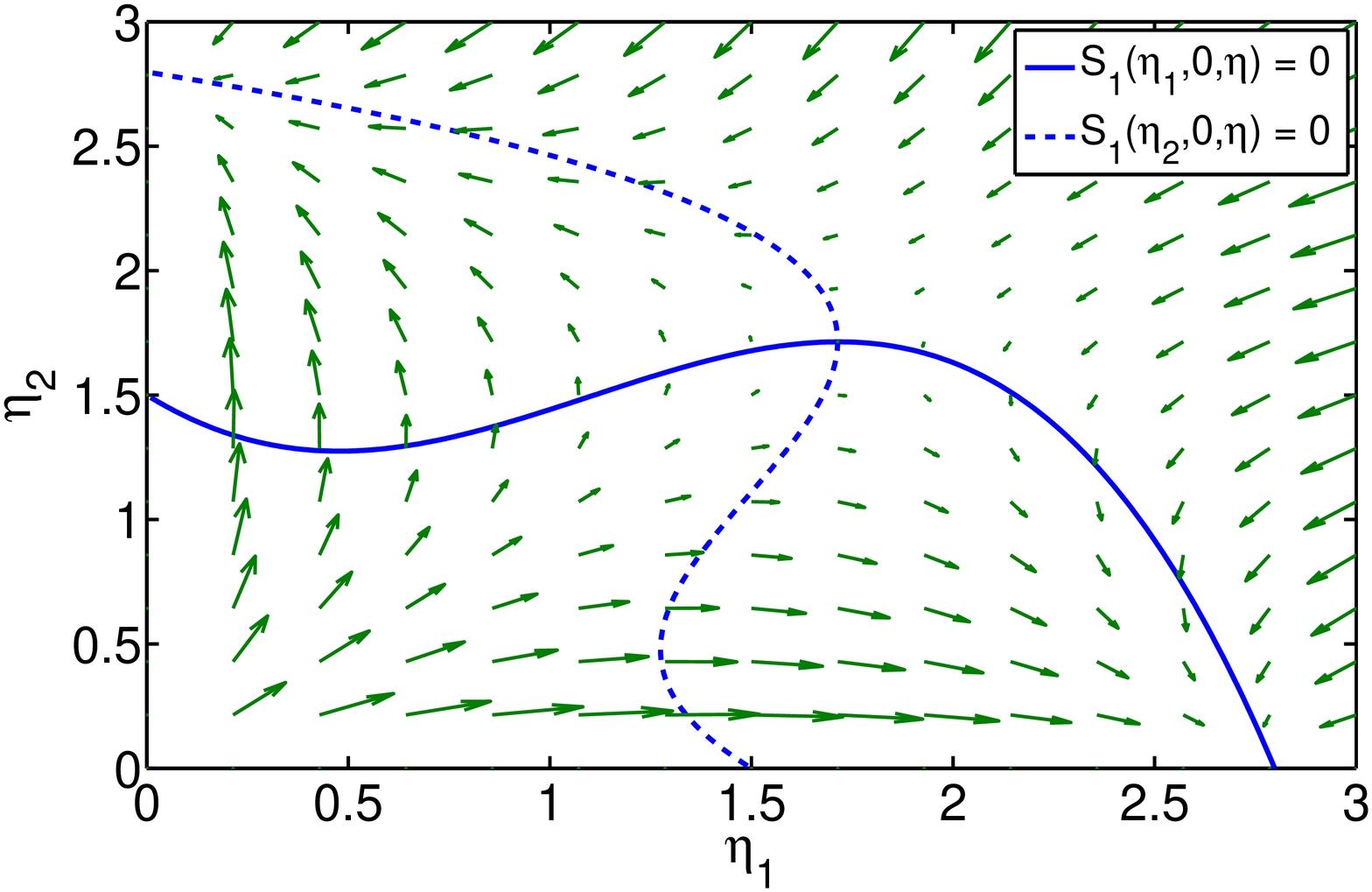}
    \includegraphics[width=2.5in]{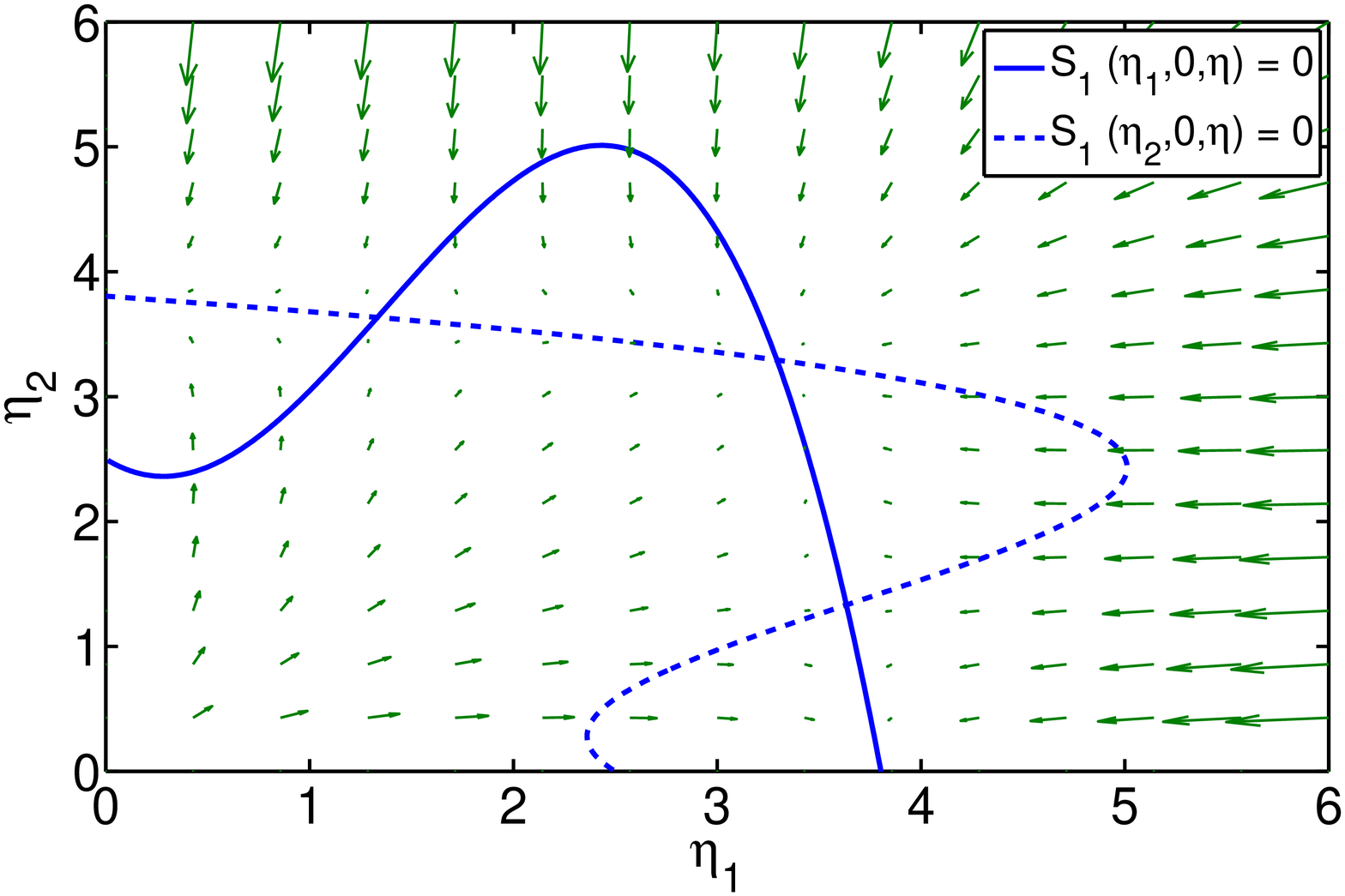}
    \caption{(Color online) Typical phase planes for $\frac{d\eta_1}{dz}$ and
    $\frac{d\eta_2}{dz}$ with $V_1 = V_2 = 0$ for the unstable
    two soliton case, $g = 0.4$ (top) and the stable two soliton case, $g = 0.6$ (bottom).}
    \label{pplane}
\end{figure}

In Eqs. (\ref{evoeqn}) the terms $S_1$ and $S_2$ are the contributions from gain,
loss and filtering, while the other terms come from the tail interactions.
Considering just the effects of gain, loss and filtering, the velocities have a
single  stable equilibrium $V_1 = V_2 = 0$ for any positive value of $\eta_1$ and
$\eta_2$ (negative values of $\eta_1$ or $\eta_2$ are only a phase shift from their
positive counter parts), so we look at the dynamics of $\eta_1$ and $\eta_2$ for
$V_1$  and $V_2$ at this equilibrium.  Fig. \ref{pplane} shows the phase plane for
gain values above and below the threshold for stable two pulse solutions.  In both
cases there exists a stable equilibrium at $\eta_k=0$ for $k=1$ or $2$ which
amounts to a reduction to the single soliton case and an unstable equilibrium when
both $\eta_1=\eta_2= 0$ for any $g>l$.  Also, in both cases we see an equilibrium
at $\eta_1 = \eta_2= \eta^*$, however, for $g<g_c$ it is found to be unstable and
for $g>g_c$, it is stable; $g_c \approx 0.45.$

In Fig. \ref{pplane} typical cases are shown. Here $g=0.4$ (top) is unstable and
$g=0.6$ (bottom) is stable. The $\eta^*$ found from perturbation theory agrees with
the height found numerically to several decimal places.  We also note that there
are two other two equilibria found for $g=0.6$ which are unstable.

\begin{figure}[!htbp]
    \centering
    \includegraphics[width=2.5in]{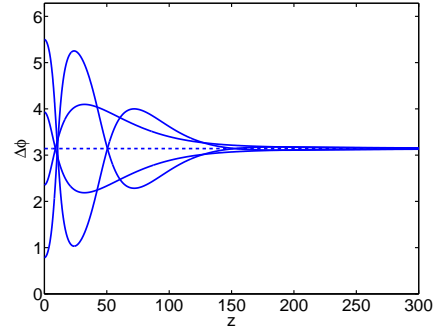}
    \caption{(Color online) Evolution of the phase for $\eta_{10} = \eta_{20} = 3.3$,
    $\Delta\xi_0 = 3.0$ ($\Delta\xi_0/\alpha \approx 6 $) and several choices of initial
    phase difference.  Here $g =0.6$. $\Delta\phi=\pi$ is shown with a dashed line.}
    \label{phases}
\end{figure}

When the soliton interaction terms are taken into consideration, $S_1$ and $S_2$
are found to still be the dominant terms for sufficiently large $\Delta \xi,$ hence
$\eta_k \rightarrow \eta^*$ and $V_k \rightarrow 0$ for $k=1,2$. However, the small
contributions from the interaction terms can cause small decaying oscillations for
large $z.$  Solving the perturbation theory derived system, we find a stable
equilibrium at $\Delta\phi = \pi$ for any $\eta$ as well as an unstable equilibrium
for $\Delta\phi=0$ and $\Delta\eta=0$.  See typical cases in Fig. \ref{phases}. The
peak separation is further characterized by the phase difference; the pulses
attract each other when $-\pi/2 \leq \Delta \phi<\pi/2$ and repel when
$\pi/2<\Delta \phi \leq 3\pi/2$.  This is seen from inspecting the sign of $\Delta
V_t = - 8 \bar{\eta}^3 \cos (\Delta\phi) \exp(\bar{\eta} \Delta\xi)$ for two
solitons with zero velocity. Thus, for initial phase difference $\Delta\phi_0=0$
the pulses will eventually collide and combine into a single pulse, while for
$\Delta\phi_0 \neq 0$ the phase difference will evolve to $\pi$ and after some
initial oscillations the pulses will repel.

\begin{figure}[!htbp]
    \centering
    \includegraphics[width=2.5in]{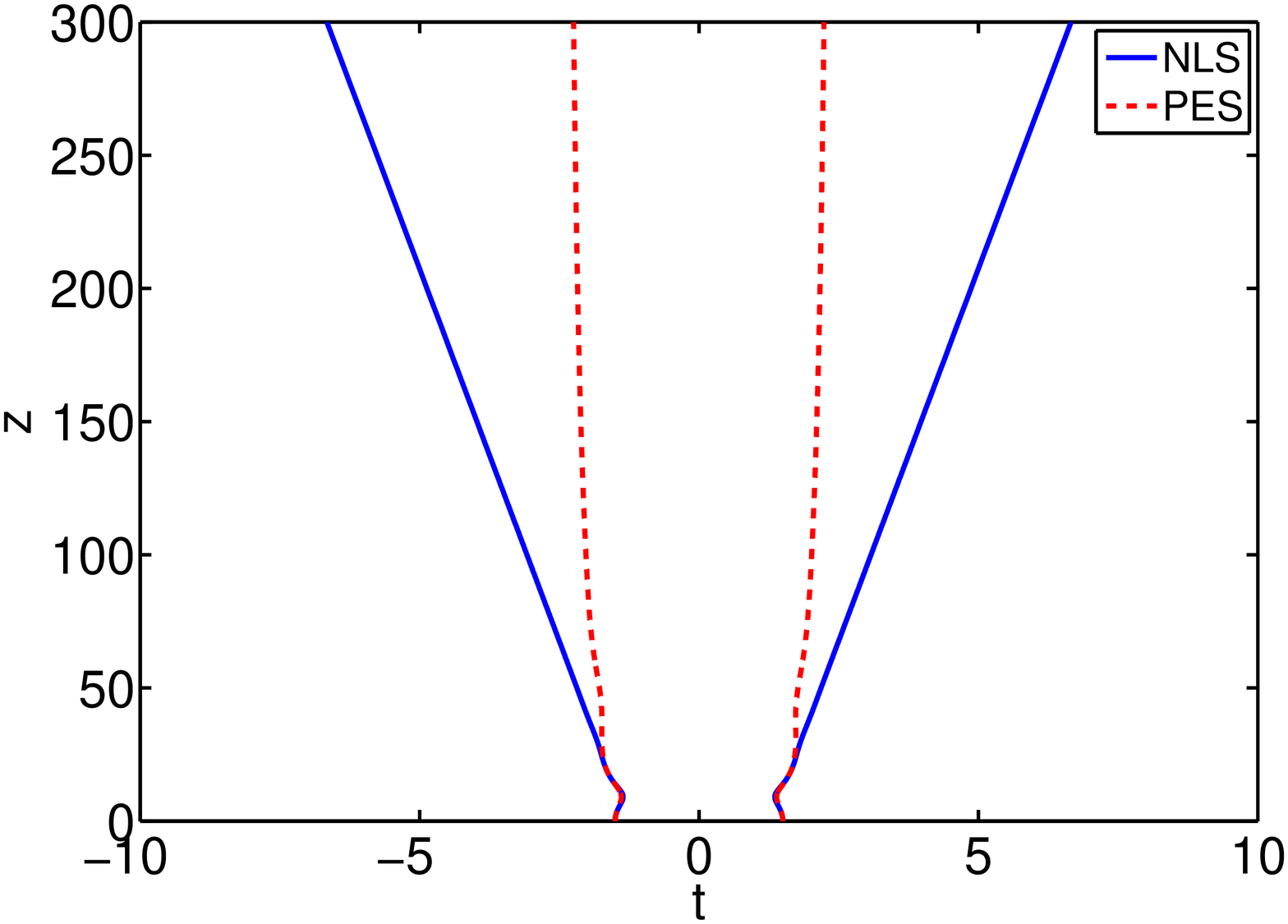}
    \includegraphics[width=2.5in]{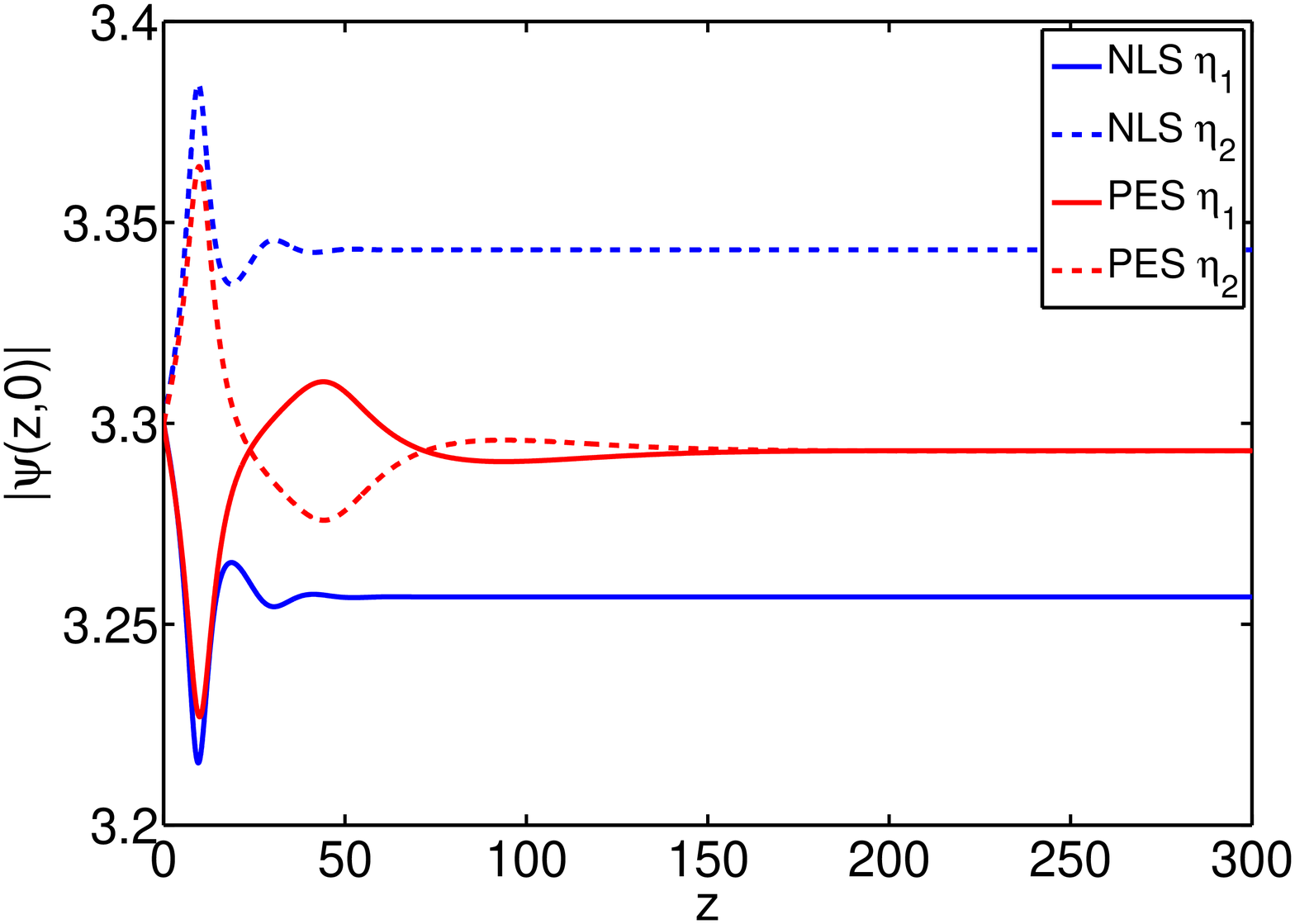}
    \caption{(Color online) Comparison of NLS to PES; initially the solitons have $\Delta \phi=\pi/4$. For
    Eq. (\ref{PES}) the resulting pulses have phase difference $\Delta \phi = \pi$.  Here $g = 0.6$.}
    \label{NLSvPES}
\end{figure}

There is not a non-trivial equilibrium for the full Eqs. (\ref{evoeqn}), so the
``effective high order soliton state found numerically is not a ``true'' bound state.  We will
show, however, that $\Delta \xi(z) = O(\log(z))$ so once the solitons are a certain
distance $d^*$ apart they will be moving too slowly to see or measure. By
comparison, for the classical NLS equation $\Delta \xi(z) = O(z)$ \cite{yijianke}.  This
difference is illustrated in Fig. \ref{NLSvPES} for $\Delta \phi_0=\pi/4$.  These
``effective" bound states emerge naturally from the interaction of pulses for most
initial conditions making them attractors of Eq. (\ref{PES}).

Using the fact that $\eta_1 = \eta_2  \rightarrow \eta^*$ and  $\Delta\phi
\rightarrow \pi$ as $z \rightarrow \infty$ we can  derive the asymptotic behavior
of $\Delta\xi$.  First we take the difference of $\frac{dV_2}{dz}$ and
$\frac{dV_1}{dz}$ and the derivative of $\frac{d \Delta\xi }{dz} $ giving us
\begin{align*}
\frac{d \Delta V}{dz}&= -\tau \left( \frac{4}{3} \frac{\eta^{*2}}{ 4 \eta^*+1}
\right) \Delta V  + 8\eta^{*3}
e^{-\eta^* \Delta\xi}\\
\frac{d^2  \Delta\xi}{dz^2} &= \frac{d \Delta V}{dz}
\end{align*}
which may now be combined to form a second order differential equation for
$\Delta\xi$
\begin{eqnarray*}
3\frac{d^2  }{dz^2}\Delta\xi =  -\tau \left( \frac{4}{3} \frac{\eta^{*2}}{ 4 \eta^*+1} \right) \frac{d  }{dz}\Delta\xi
+  8\eta^{*3} e^{-\eta^* \Delta\xi} = 0
\end{eqnarray*}
Since the evolution of $\Delta\xi$ is seen from numerical simulations to be
evolving slowly in $z$ we expect that $\frac{d^2 }{dz^2}\Delta\xi \ll \frac{d
}{dz}\Delta\xi$ and so $\frac{d^2  }{dz^2}\Delta\xi$ can be neglected as being a
higher order term. This leaves a first order equation which we solve to get the
long term behavior of $\Delta\xi$ to be
\begin{eqnarray*}
\Delta\xi &\rightarrow&  \frac{1}{\eta^*} \log \left[\frac{6\eta^* (1+4\eta^*)}{\tau}  z \right]
\end{eqnarray*}
In fact it is found numerically that the divergence is considerably less  than
$\log z$. Hence we call these higher states ``effectively bound'' because in
practice the laser propagation distance is bounded.

\begin{figure}[!htbp]
    \centering
    \includegraphics[width=3.5in]{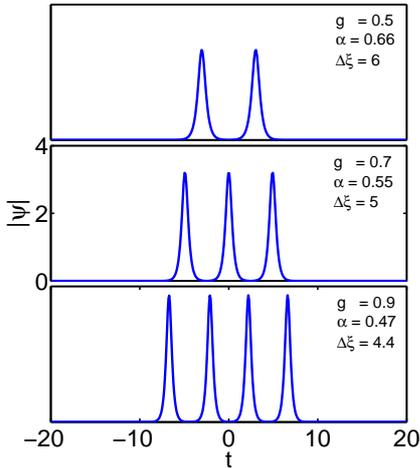}
    \caption{(Color online) Multiple high-order bound state soliton solutions of
    the anomalously dispersive PES model.}
    \label{all.fig}
\end{figure}

As mentioned above, by increasing the gain parameter additional pulses may be
supported. In this way we can find three and four soliton states that emerge from
unit gaussian initial states as shown in Fig. \ref{all.fig}.  In all cases pulses
of height $\eta^*$ were taken at an initial separation of $\Delta\xi = 9\alpha$ and
evolved to $z=1000$ with less then $0.1$ change in peak positions.

The depicted solitons are all in phase, however such higher order states can be
found for any choice of (initial) pair wise phase differences. Interestingly, for
no interactions to occur the critical separation distance between the solitons is
again found to be $d^*\approx9\alpha$ for the three and four soliton states. For
$\Delta \xi<d^*$ the interactions of three and four solitons are more varied, but
essentially result in either evolving to a lower state through the merging of
pulses or pulses repelling each other beyond the critical distance $d^*$. This
behavior agrees with what was found for the two soliton state.

\section{Normal Higher Order Solitons}

\begin{figure}[!htbp]
    \centering
    \includegraphics[width=2.5in]{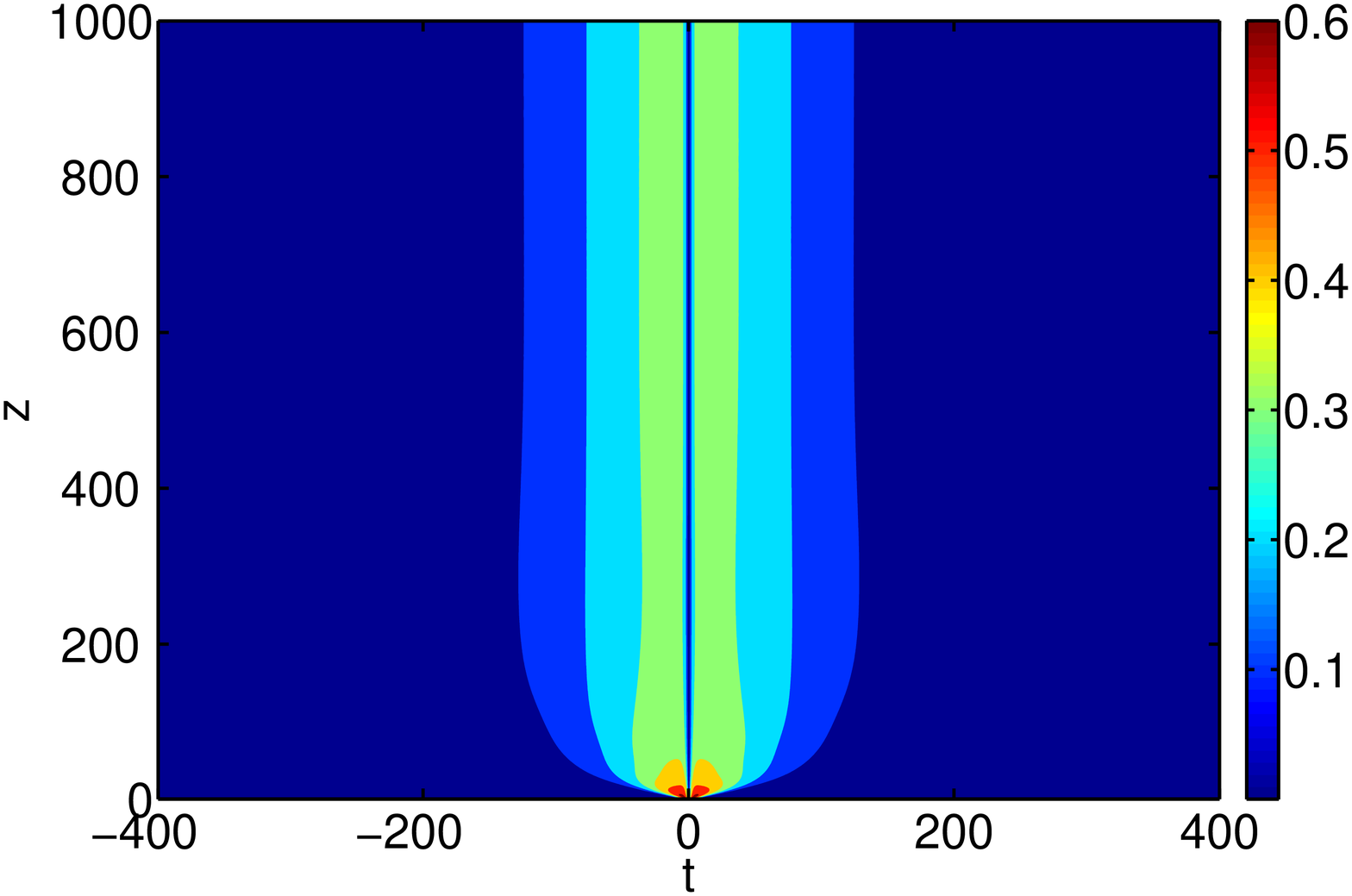}
    \includegraphics[width=2.5in]{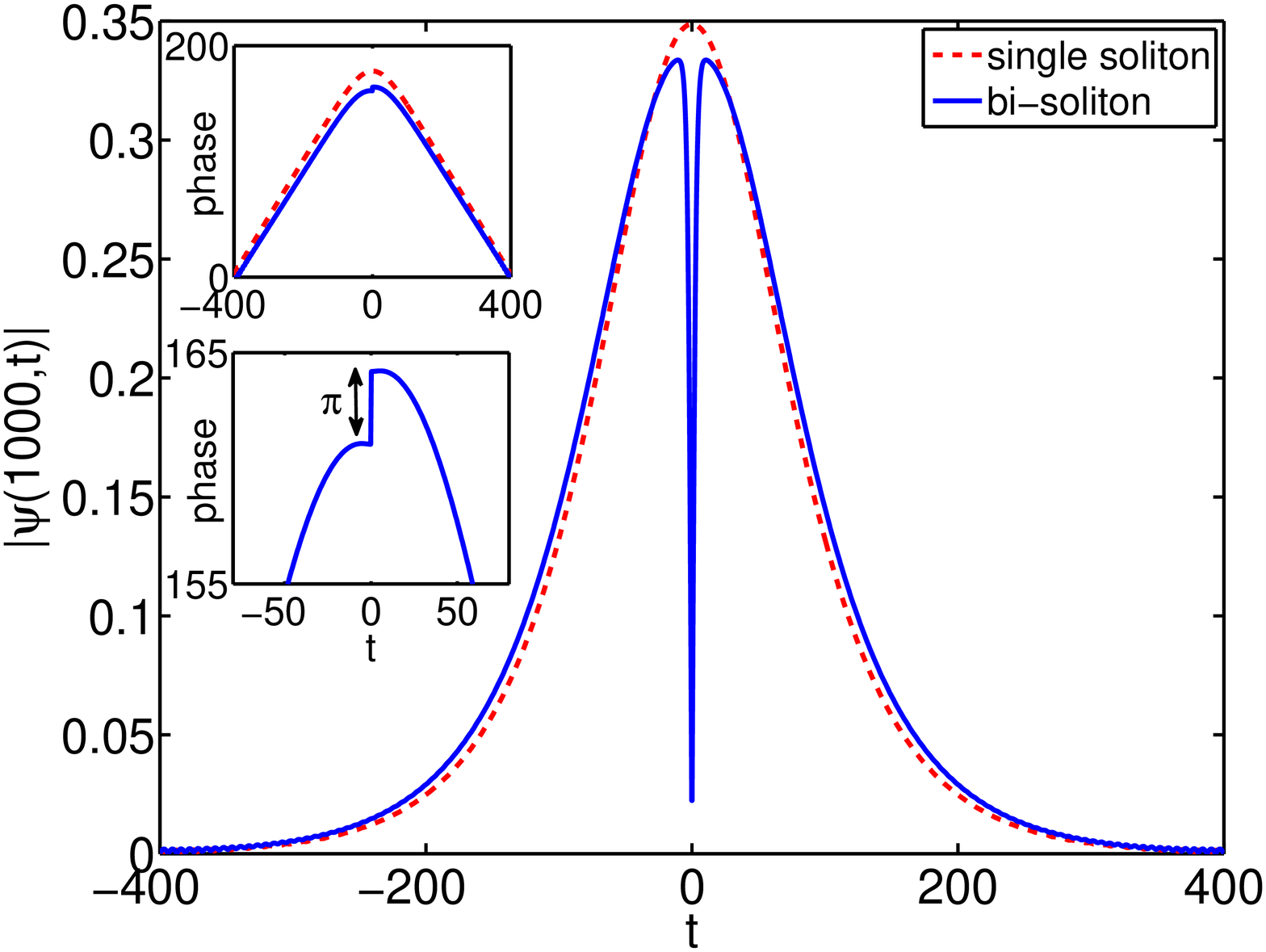}
    \caption{(Color online) Evolution of the anti-symmetric soliton (top) and the anti-symmetric
    (bi-soliton) superimposed with the relative single soliton (bottom) at $z=1000$. Here $g=1.5$.}
    \label{normal}
\end{figure}

Next we briefly mention some results for the constant normal dispersive case:
$d_0=-1<0$.  As mentioned above, individual pulses of  Eq. (\ref{PES}) in the
normal regime exhibit strong chirp and cannot be identified as the solutions of the
``unperturbed'' NLS equation. Indeed, the classical NLS equation does not exhibit
decaying ``bright'' soliton solutions in the normal regime. If we begin with an
initial gaussian, $\psi(0,t)=\exp(-t^2)$, the evolution mode-locks into a
fundamental soliton state; see Fig. \ref{single.fig} bottom two figures. These
figures clearly exhibit the mode-locking evolution and the significant chirp of the
pulse. On the other hand we can obtain a higher-order anti-symmetric soliton, i.e.
an anti-symmetric bi-soliton, one which has its peaks amplitudes differing in phase
by $\pi$. Such a state can be obtained if we start with an initial state of the
form $\psi(0,t)=t\exp(-t^2)$ (i.e. a Gauss-Hermite polynomial). The evolution
results in an anti-symmetric bi-soliton and is shown in Fig. \ref{normal} along
with a comparison to the profile of the single soliton. This is a true bound state.
Furthermore, the results of our study finding anti-symmetric solitons in the normal
regime are consistent with recent experimental observations \cite{chong2}.

\begin{figure}[!htbp]
    \centering
    \includegraphics[width=2.5in]{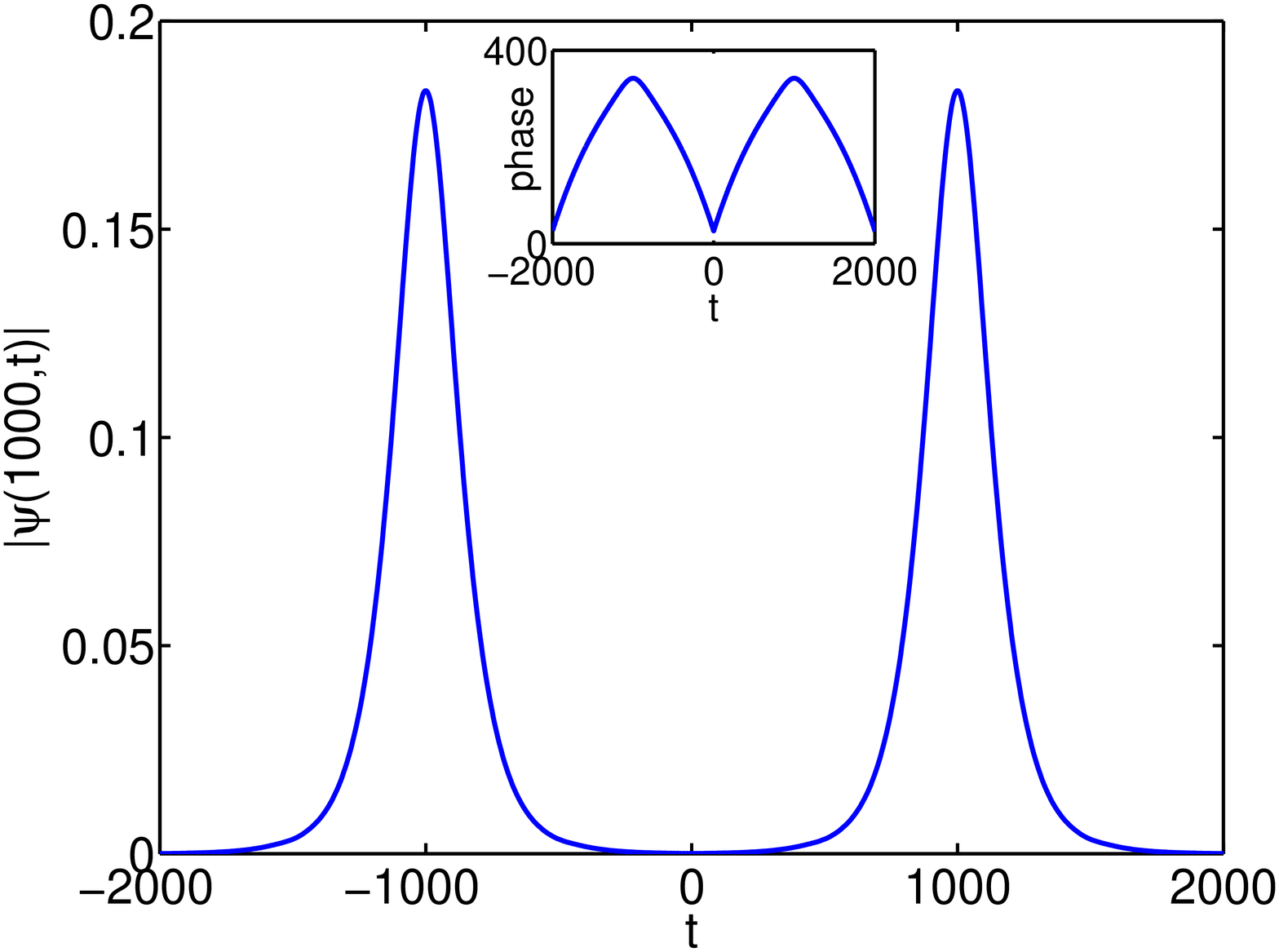}
    \caption{(Color online) Symmetric two soliton state of the normal regime with in phase pulses. The phase
    structure is depicted in the inset. Here $ g=1.5.$}
    \label{2normal}
\end{figure}

It is interesting that the normal regime also exhibits higher soliton states when
two general initial pulses (e.g. unit gaussians) are taken sufficiently far apart.
The resulting pulses, shown in Fig. \ref{2normal}, individually have a similar
shape to the single soliton of the the normal regime with lower individual
energies. These pulses if initially far enough apart can have independent chirps
which may be out of phase by an arbitrary constant. We again find that $d^*\approx
9 \alpha$ is a good estimate for the required pulse separation just as in the
constant anomalous dispersive case! These pulses are ``effective bound states'' in
that after a long distance they separate very slowly (too slowly to measure).

Additional higher order $3,4,...$ soliton states can also be found. We will not go
into further details here.

\section{Dispersion managed systems}\label{dm.section}

Standard dispersion managed (DM) solitons, like their constant dispersive
counterparts, are obtained from the PES model over a wide range of anomalous
dispersion \cite{horikis}. This is important since recent mode-locked laser
experiments have been conducted in dispersion-managed regimes
\cite{Ildaypsat,chong2,Kartner}. In correspondence with Ti:Sapphire DM laser
systems we allow the dispersion to vary in $z$ as well as introducing nonlinear
management in the form of the function $n(z)$ multiplying the nonlinear term.
Equation  (\ref{PES}), is used but we change the symbol of the envelope from
$\psi=\psi(z,t)$ to $u=u(z,t)$ to distinguish the two cases (constant dispersion
and DM, respectively). The effect of dispersion management is obtained by splitting
the dispersion $d(z)$ into two components $d(z) = d_0 + \Delta(z/z_\alpha)/z_\alpha
$ where $0<z_\alpha<<1$ is the dispersion-map period. Hence $d(z)$ is large and
periodic. If $d(z)$ was only $O(1)$ then the multi-scale averaging method employed
below would result in the constant dispersive case discussed earlier. The path
averaged dispersion is $d_0$ and $\Delta(z/z_\alpha)$ is a rapidly varying function
with zero average which we define as follows: $\Delta (\zeta) = \{-\Delta_1, 0
<\zeta<1/2; \Delta_2, 1/2<\zeta<1\}$. Here we consider the case of positive average
dispersion $(d_0>0)$. We define the map strength $s = \Delta_1/2$ which gives a
measure of the variability of dispersion around the average.  We take  $d_0 = 1,
z_a=0.1$ and vary $s$. The effect of the nonlinear management is to turn the
nonlinearity ``off and on", i.e. $n(\zeta)= \{0, 0<\zeta<1/2; 1, 1/2<\zeta<1\}$.
For example, in a Ti:Sapphire laser the nonlinearity is negligible in the anomalous
regime where one has a prism pair that compensates for the normal dispersion in the
crystal. The averaged equation is derived using the method of multiple scales and
perturbation theory \cite{biondini}. The variation in dispersion occurs on the
short distance scale $\zeta = z/z_\alpha$ and the pulse envelope evolves on the
long scale $Z=z$.

The method proceeds by expanding $u$ in powers of $z_\alpha$:
 $$u(\zeta,Z,t) = u^{(0)}(\zeta,Z,t)
+ z_\alpha u^{(1)}(\zeta,Z,t) + O(z_\alpha ^2)$$ and substituting this into Eq.
(\ref{PES}) to obtain a series of equations by relating terms by powers of
$z_\alpha$. At $O(z_\alpha^{-1})$
\begin{eqnarray*}
i \frac{\partial u^{(0)}}{\partial \zeta} + \frac{\Delta(\zeta)}{2} \frac{\partial^2  u^{(0)}}{\partial t^2} = 0
\end{eqnarray*}
which can be solved using Fourier transforms to arrive at
\begin{eqnarray*}
\hat{u}^{(0)} ( \zeta, Z, \omega) = \hat{U}_0 (Z, \omega)\exp \left[ - i \frac{\omega}{2} C(\zeta) \right]
\end{eqnarray*}
where $C(\zeta)= \int_0^{\zeta} \Delta(\zeta')d\zeta'$, $\hat{U}_0(Z,\omega) =
\hat{u}^{(0)}(\zeta = 0 , Z, \omega)$, and the Fourier transform pair is defined as
\begin{align*}
\hat{u}(\omega)=\mathcal{F}\{ u(t) \}&=\intl u(t) e^{i\omega t}dt\\
u(t)=\mathcal{F}^{-1}\{ \hat{u}(\omega) \}&=\frac{1}{2\pi}\intl \hat{u}(\omega) e^{-i\omega t}d\omega
\end{align*}

Thus $\hat{u}_0$ separates into a slowly evolving envelope $\hat{U}_0$ and fast
oscillations due to changes in the local dispersion. The equation for $\hat{U}_0$
is obtained by imposing secularity conditions on the $O(1)$ terms,
\begin{align}
i \frac{\partial \hat{U}_0}{\partial Z} - \frac{d_0}{2}\omega^2 \hat{U}_0 &+ \disp\int_0^1 \exp \left[ - i
\frac{\omega}{2} C(\zeta)\right] (\mathcal{F} \{|u^{(0)}|^2 u{(0)} \} \nonumber\\&- \mathcal{F} \{ F[u^{(0)} ]
\})d\zeta = 0,
\label{DMPES}
\end{align}
This is the averaged, or mean-field equation which we solve to find the pulse
dynamics. In the case where $F=0$ this is known as the dispersion managed NLS or
DMNLS equation \cite{biondini}.

\begin{figure}[!htbp]
    \centering
    \includegraphics[width=2.5in]{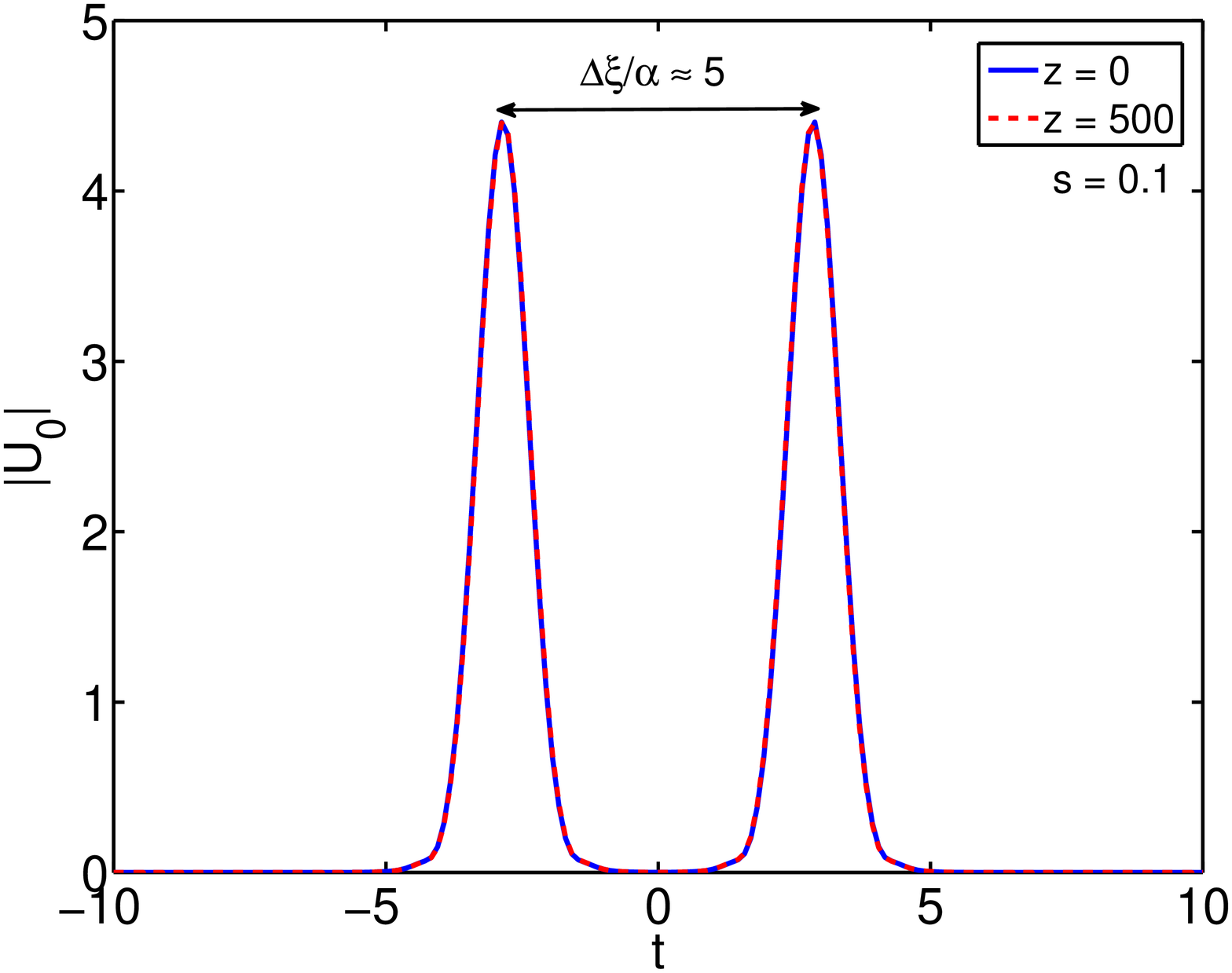}
    \includegraphics[width=2.5in]{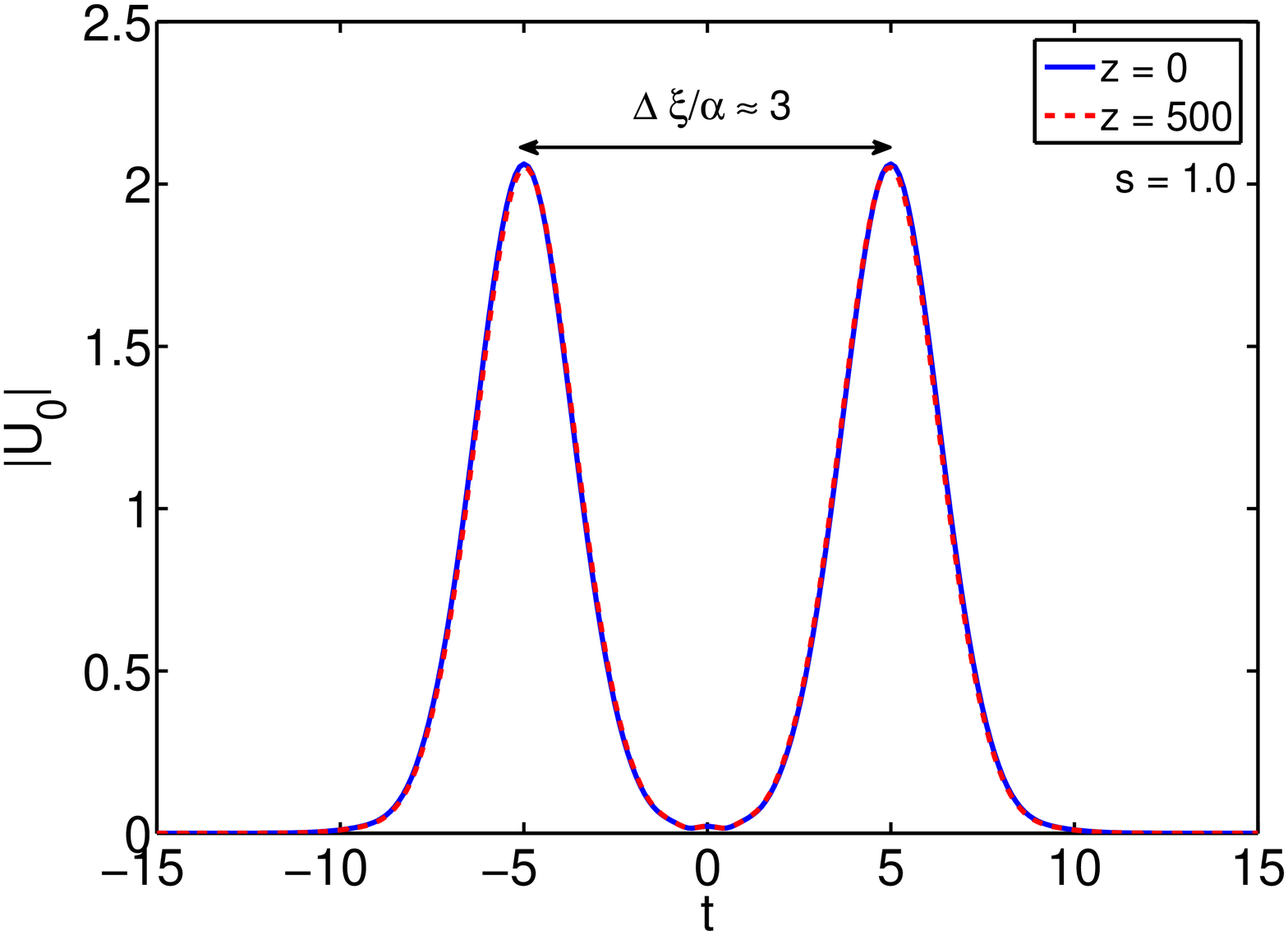}
    \caption{(Color online) ``Effective" bound state for two solitons in DM system for $s=0.$ (top)
    and $s=1.0$ (bottom).  Here $g =0.6$.}
    \label{DMfig1}
\end{figure}

\begin{figure}[!htbp]
    \centering
    \includegraphics[width=2.5in]{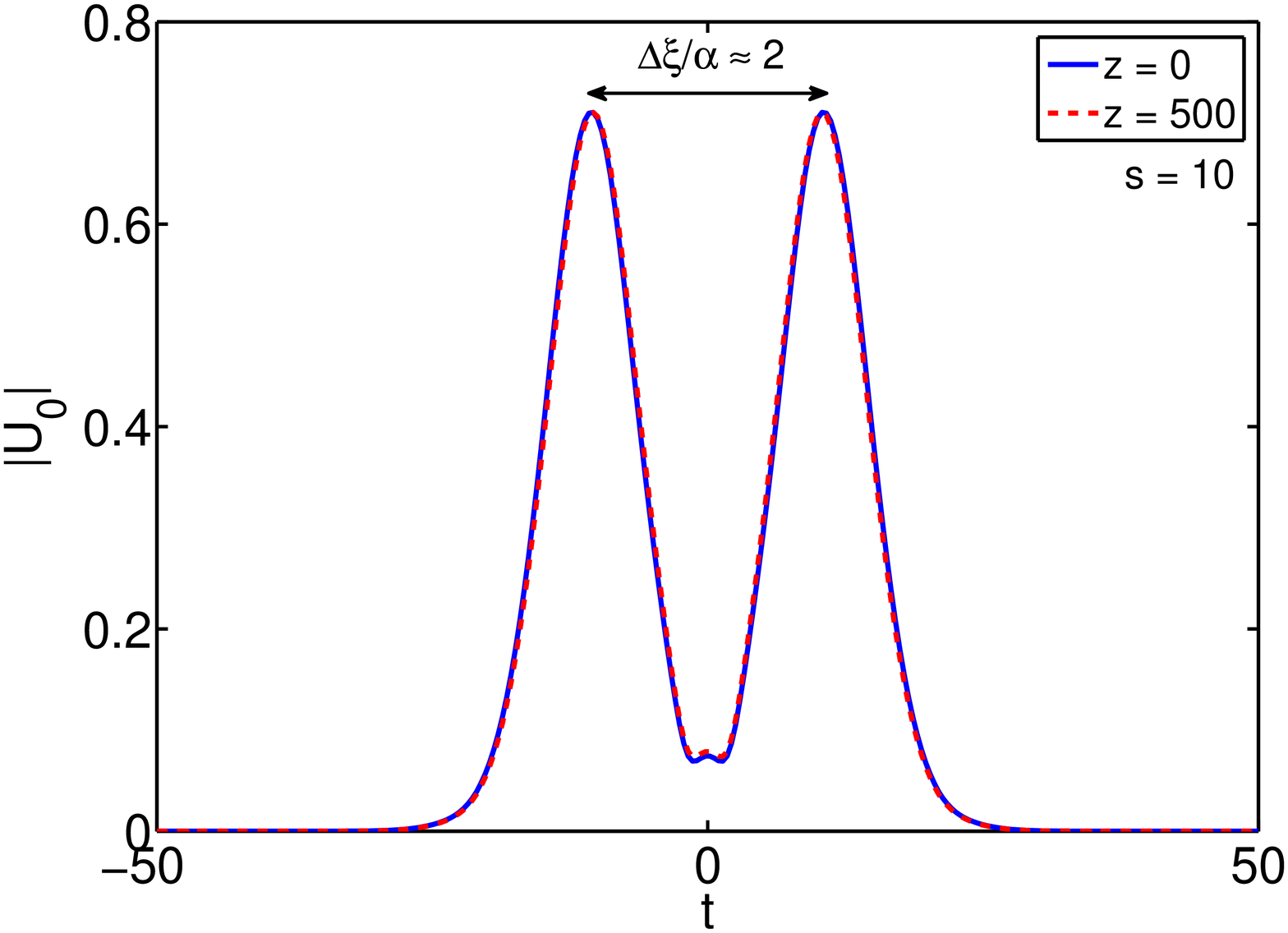}
    \includegraphics[width=2.5in]{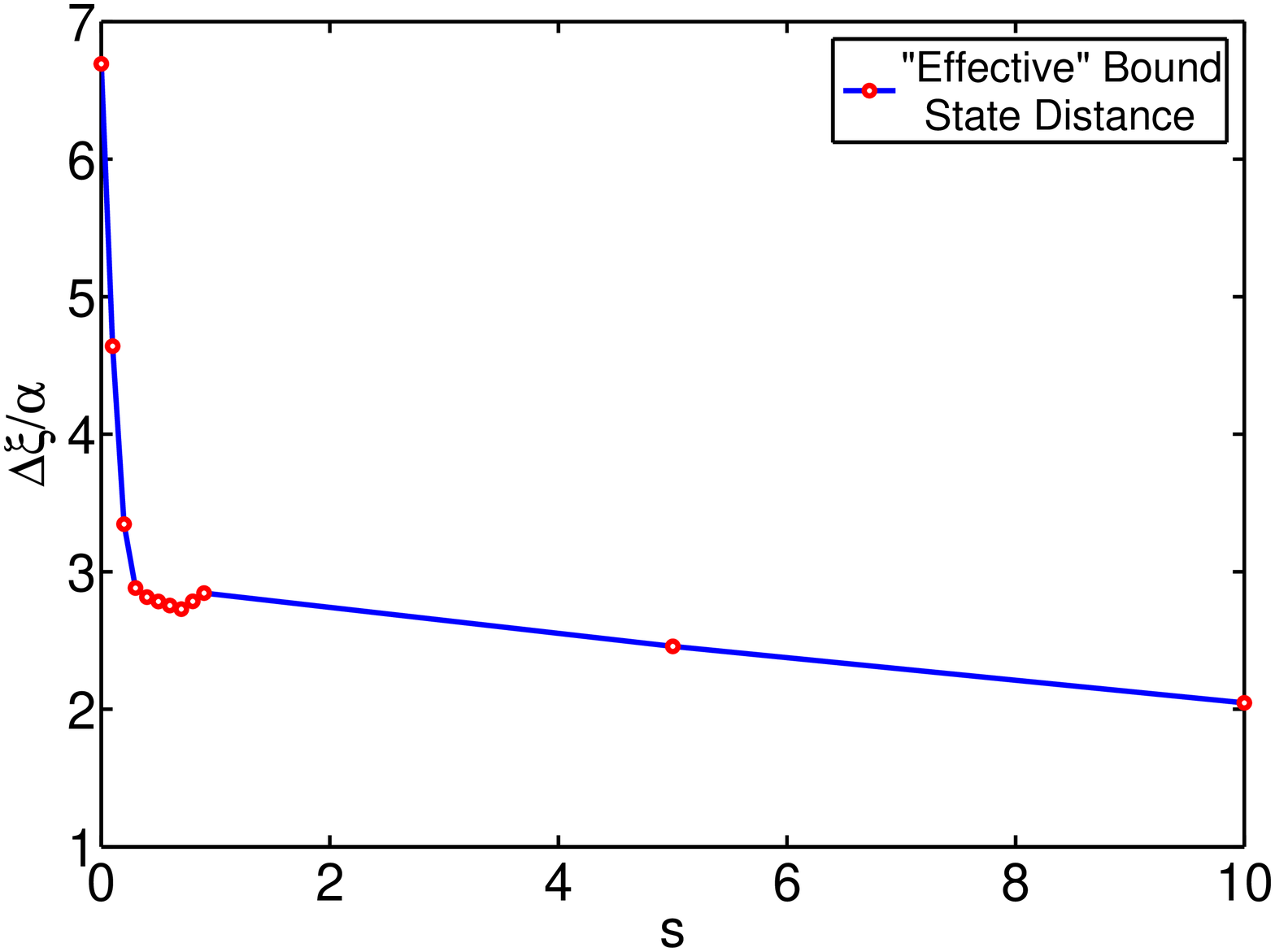}
    \caption{(Color online) ``Effective" bound state for two solitons in DM
    system for $s=10$ (top). Numerically found relation between the map
    strength $s$ and the minimum initial distance $\Delta\xi_0/\alpha$ for no interactions to
    be seen (bottom) after $z=500$ units.  Here $g = 0.6$.}
    \label{DMfig2}
\end{figure}

The method of spectral renormalization \cite{spectral1} can be employed to find
single mode-locked DM solitons for Eq. (\ref{DMPES}) \cite{horikis}.  Here we
initially superimpose two such DM soliton pulses at varying peak separations with
phase difference $\Delta\phi = 0$ and let them evolve to find two soliton
``effective" bound states.  As a criterion for an ``effective" bound state, we
require the the peak separation differ less than $0.05\alpha$ after evolving $500$
units in $z$.  Typical examples with comparison of initial vs. final states are
depicted in Figs. \ref{DMfig1} and \ref{DMfig2}.  Much like the single soliton case
the individual pulses are well approximated by the solutions of the unperturbed
DMNLS equation.  We also note that due to the nonlocality of the equation, the
individual pulses have a smaller energy than the single pulse for the same
parameters, as was in the case of constant dispersion.

As is indicated in the figure, the minimum initial distance $d^* \approx 9\alpha$
no longer holds for the dispersion (and nonlinearly) managed PES equation (DMPES).
In the nonlinear managed system we find that the critical distance $d^* \approx
7\alpha$ for $s=0$, i.e. constant dispersion, and $d^*$ depends on the map strength
$s$ for $s>0$.  Since this change is much more dramatic between $s=0$ and $s=1$
than $s=1$ and $s=10$ we investigate this region more thoroughly. In Fig.
\ref{DMfig2} the value of $\Delta \xi / \alpha$ found for the pulses to be
``effectively" bound are plotted for varying map strengths. The general trend is
for the needed separation to decrease as $s$ increases, however, as can be seen
between $s=0.1$ and $s=1.0$ where more $s$ values were tested this is not a
monotonically decreasing process.

For more information on the normal ($d_0<0$) DM case we refer the reader to Ref.
\cite{AHnormal}.

\section{Anti-Symmetric Bi-Solitons for DM Systems}

\begin{figure}[!htbp]
    \centering
    \includegraphics[width=2.5in]{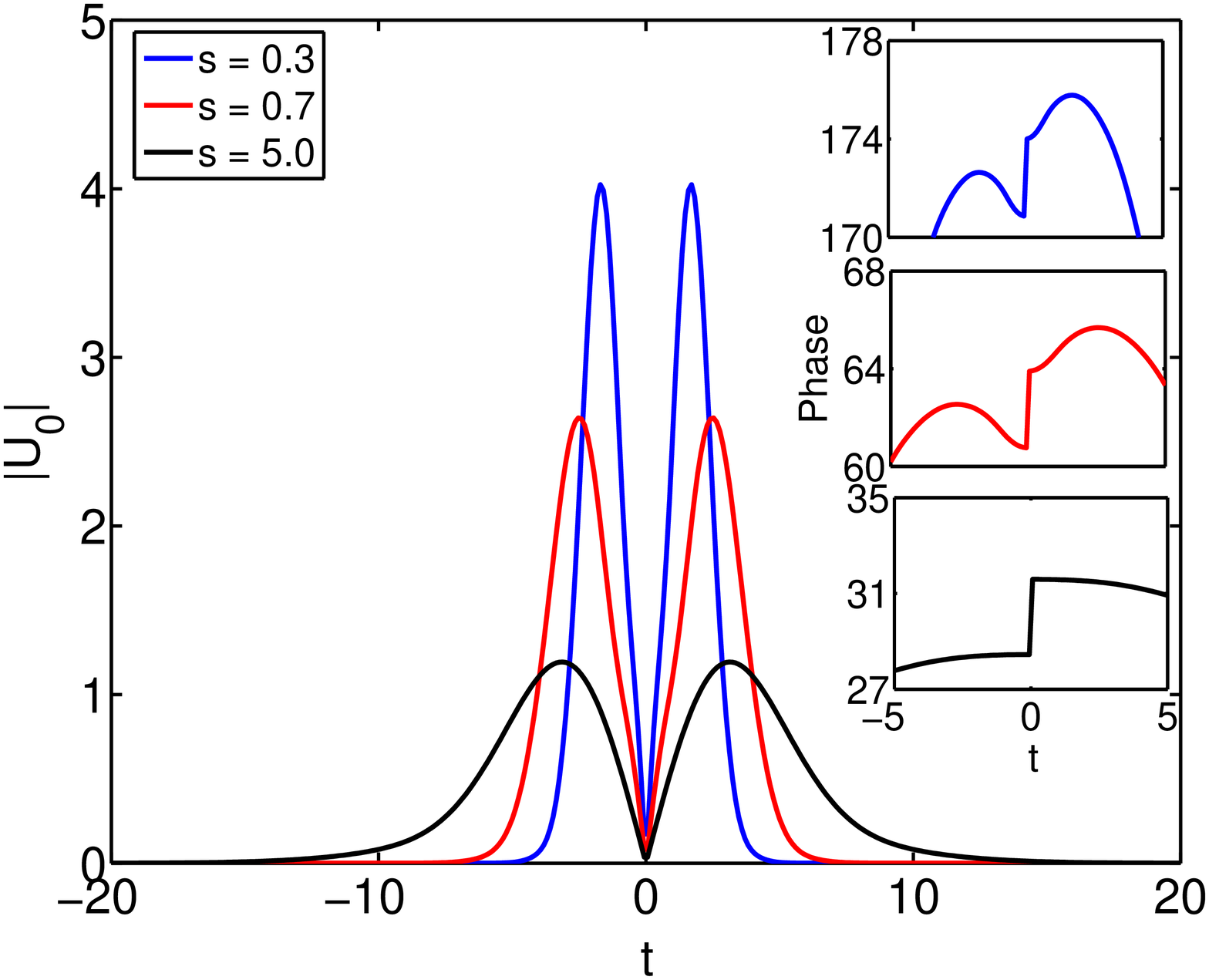}
    \caption{(Color online) Bi-Soltion states found for $s = 0.3, 0.7$ and 5.0. Here $g = 0.6$. The relative
    phases and the $\pi$ jumps at the origin are shown in the insets.}
    \label{BiSolitonDM}
\end{figure}

We now superimpose two net anomalous DM solitons with a $\pi$ phase difference. For
$s<s_* \approx 0.25$ pulses repel  each other as was the case for constant
anomalous dispersion; we note that for the above values of the map strength both of
the local dispersions being used ($d_0 + \Delta_1/z_a$ and $d_0 + \Delta_2/z_a$)
are in the anomalous regime.  For $s>s_*$,  pulses which are taken close enough
together, i.e. the distance between peak values, $d<d_* \approx 2.5,$  are found to
lock into a bi-soltion state.  Examples are given in Fig. \ref{BiSolitonDM}. Pulses
taken further apart:  $d>d_*$ are found to repel. These anti-symmetric bi-solitons
are the mode-locked (due to gain-loss) analog of what was obtained in the case of
pure DM systems  without gain loss \cite{maruta2,hirooka}.

\section{Summary}

To conclude, we investigated Eq. (\ref{PES}) and found a large class of localized
solutions including: mode-locked solitons in both the constant anomalous and normal
regimes, high-order  solitons in the constant anomalous regime and anti-symmetric
bi-solitons in the constant normal regime. These results are consistent with
experimental observations of higher-order solitons in the anomalous and bi-solitons
in the normal dispersive regimes. The dispersion and nonlinear managed system was
also investigated. Here in the  averaged anomalous regime single and higher-order
soliton pulses were obtained, including anti-symmetric bi-solitons in the net
anomalous regime. For the constant dispersion case,  it is found that when
individual pulses are initially separated by $d^*\approx 9 \alpha$ where $\alpha$
is the width of the individual pulse the result is a soliton string. For the DM
system the results indicate that the high-order soliton strings  in the DM case can
exist in much closer proximity to each other.

\section*{Acknowledgments}

This research was partially supported by the U.S. Air Force Office of Scientific
Research, under  grant FA9550-09-1-0250; by the National Science Foundation, under
grant DMS-0505352. \\ We appreciate many valuable discussions with Dr. Y. Zhu.

\bibliographystyle{elsarticle-num}

\end{document}